\documentclass[acmsmall,screen,nonacm]{acmart}

\acmJournal{PACMPL}
\acmVolume{1}
\acmNumber{PLDI}
\acmArticle{1}
\acmYear{2026}
\acmMonth{6}
\setcopyright{cc}
\setcctype{by}

\usepackage[T1]{fontenc}

\usepackage{appendix}
\usepackage{float}
\usepackage{mathpartir}
\usepackage{siunitx}
\usepackage{footnote}
\usepackage{amsmath}
\usepackage{amsfonts}
\usepackage{mathtools}
\usepackage{proof}
\usepackage{latexsym}
\usepackage{stmaryrd}
\usepackage{mdframed}
\usepackage{enumitem}
\usepackage{graphicx}
\usepackage{subcaption}
\usepackage[dvipsnames]{xcolor}
\usepackage{xspace}
\usepackage{booktabs}
\usepackage{array}
\usepackage{multicol}
\usepackage{multirow}
\usepackage{colortbl}
\usepackage{makecell}
\usepackage{tablefootnote}
\usepackage{algorithm}
\usepackage[noend]{algpseudocode}
\usepackage{listings}
\usepackage{xcolor}
\lstdefinestyle{pythonstyle}{
  language=Python,
  basicstyle=\ttfamily\scriptsize,
  commentstyle=\color[RGB]{106,153,85}\ttfamily\scriptsize,
  keywordstyle=\color[RGB]{86,156,214}\bfseries,
  stringstyle=\color[RGB]{206,145,120},
  breaklines=true,
  frame=none,
  numbers=none,
  numberstyle=\ttfamily\scriptsize,
  numbersep=6pt,
  showstringspaces=false,
}
\usepackage{fvextra}
\input{pygments_style}
\usepackage{listings}
\usepackage{courier}
\usepackage{refcount}
\usepackage{balance}
\usepackage{tabularx}
\usepackage{multicol}
\usepackage[most]{tcolorbox}
\usepackage{fontawesome5}
\tcbuselibrary{skins,breakable}

\usepackage[table]{xcolor}
\usepackage{colortbl}

\newcommand{\code}[1]{\texttt{#1}}

\lstdefinelanguage{lark}{
  morekeywords={constraint,expression,term,factor,property_access,
    date_constructor,period_constructor,comparison_expr},
  sensitive=true,
  morecomment=[l]{\#},
}

\definecolor{extgray}{RGB}{235,235,235}

\newcommand{\datesat}{\textsc{DateSAT}\xspace}
\newcommand{\dataset}{\textsc{DateSATBench}\xspace}
\newcommand{\totalConstraints}{800\xspace}

\newcommand{\naive}{\textsc{Simple}\xspace}

\newcommand{\epoch}{\textsc{Epoch}-based\xspace}
\newcommand{\hybrid}{\textsc{Hybrid}\xspace}
\newcommand{\hybridymd}{\textsc{Hybrid}\textsubscript{\textit{init\_ymd}}\xspace}
\newcommand{\hybridepoch}{\textsc{Hybrid}\textsubscript{\textit{init\_epoch}}\xspace}

\newcommand{\ab}{\textsc{Alpha-Beta}\xspace}

\newcommand{\Date}[1]{|#1|}
\newcommand{\Dates}{\mathcal{D}}
\newcommand{\Period}[1]{\langle #1 \rangle}
\newcommand{\Periods}{\mathcal{P}}
\newcommand{\nbdays}[1]{\mathit{dim}(#1)}
\newcommand{\plusMonths}{+_\textsc{m}}
\newcommand{\plusDays}{+_\textsc{D}}
\newcommand{\plusEither}{+_\delta}
\newcommand{\plusPeriod}{+_\Periods}

\newcommand{\true}{\mathrm{true}}
\newcommand{\false}{\mathrm{false}}

\AtEndPreamble{%
  \theoremstyle{acmplain}
  \newtheorem{property}[theorem]{Property}
}

\definecolor{subtleyellow}{HTML}{FFDE21}

\definecolor{forestgreen}{HTML}{2E6F40}

\definecolor{hypblue}{HTML}{1F4E79}
\newlist{hypotheses}{enumerate}{1}
\setlist[hypotheses]{label=\textbf{H\arabic*}, ref=H\arabic*, leftmargin=*, labelsep=0.5em,
    itemsep=1.5pt, parsep=0pt, topsep=0pt, partopsep=0pt, resume}

\definecolor{dsbA}{HTML}{EDA100}
\definecolor{dsbB}{HTML}{E87BA4}
\definecolor{dsbC}{HTML}{008300}
\definecolor{dsbD}{HTML}{6250D6}
\colorlet{grammarhighlight}{dsbA!28}
\colorlet{dsbAfill}{dsbA!28}             \colorlet{dsbAframe}{dsbAfill}
\definecolor{dsbBfill}{HTML}{EE9DBB}     \colorlet{dsbBframe}{dsbBfill}
\definecolor{dsbCfill}{HTML}{62AD62}     \colorlet{dsbCframe}{dsbC!60!black}
\definecolor{dsbDfill}{HTML}{D2CBF7}     \colorlet{dsbDframe}{dsbD!75!black}
\newcommand\dsbhl[2]{{\setlength\fboxsep{0.4pt}\setlength\fboxrule{0.3pt}\kern-0.7pt\fcolorbox{#1frame}{#1fill}{\vphantom{()}#2}\kern-0.7pt}}
\newcommand\dsbsw[1]{{\setlength\fboxsep{0pt}\setlength\fboxrule{0.3pt}\fcolorbox{#1frame}{#1fill}{\rule[-0.1ex]{0pt}{0.75em}\hspace{0.75em}}}}

\newcommand{\hypissue}[1]{\href{https://github.com/HypothesisWorks/hypothesis/issues/#1}{\##1}}
\newcommand{\hyppr}[1]{\href{https://github.com/HypothesisWorks/hypothesis/pull/#1}{\##1}}
\newcommand{\chissue}[1]{\href{https://github.com/pschanely/CrossHair/issues/#1}{\##1}}
\newcommand{\chpr}[1]{\href{https://github.com/pschanely/CrossHair/pull/#1}{\##1}}

\newcommand{\datesatrepo}{https://github.com/cmu-pasta/DateSAT}
\newcommand{\datesatbenchrepo}{https://github.com/cmu-pasta/DateSATBench}

\begin{document}

\citestyle{acmnumeric}

\title{\datesat: A Framework for Solving Date and Period Constraints}

\author{Leyi Cui}
\affiliation{%
  \institution{Carnegie Mellon University}
  \city{Pittsburgh}
  \state{Pennsylvania}
  \country{USA}
}
\email{angellc@cmu.edu}

\author{Shrey Tiwari}
\affiliation{%
  \institution{Carnegie Mellon University}
  \city{Pittsburgh}
  \state{Pennsylvania}
  \country{USA}
}
\email{shrey@cmu.edu}

\author{Rohan Padhye}
\affiliation{%
  \institution{Carnegie Mellon University}
  \city{Pittsburgh}
  \state{Pennsylvania}
  \country{USA}
}
\email{rohanpadhye@cmu.edu}

\begin{abstract}
Dates and calendar periods (i.e., days, months, years) appear frequently in tasks involving analysis of software, data, and documents. Prior research has shown that computer logic involving dates and calendrical calculations is error-prone due to tricky rules (e.g., irregularly sized months), ambiguities (e.g., scheduling one month from ``January 31st''), and edge cases (e.g., leap years). However, existing program analysis and verification tools do not provide native support for dates, making it hard to reason symbolically about calendrical arithmetic.

This paper presents \datesat, the first framework for expressing and solving satisfiability constraints involving dates and calendar periods. The paper first formalizes an input language and the semantics of date and period arithmetic. 
The paper then presents four strategies for solving \datesat{} constraints based on reductions to SMT formulas involving integers, with mechanized proofs of their correctness, and an implementation of each using Z3 as a backend.
We curate a dataset of \totalConstraints \datesat{} constraints synthesized using LLM prompting, grammar-based sampling, and mining legal documents, and present an empirical evaluation of \datesat{} solver performance, which shows that the encodings have complementary strengths and that routing each constraint to a suitable encoding solves more constraints than any single encoding.
\end{abstract}



\maketitle

\section{Introduction}
\label{sec:introduction}

\emph{``The day before yesterday I was 25 and next year I will turn 28.'' When can this be true?}~\cite{BirthdayPuzzle}

Calculations involving dates and calendrical periods make for fun newspaper-style puzzles, but they also occur commonly in various real-world applications from flight scheduling and payroll to finance and legal compliance. Such reasoning about dates is often tricky and error-prone, due to potential ambiguity (e.g., determining what is ``1 month from January 31st'') and violation of algebraic laws such as invertibility (e.g., subtracting ``1 month'' from the previous result does not necessarily return ``January 31st''). Mistakes in handling date/period logic can be costly: for example, a researcher discovered that a bug in the US~Patent and Trademark Office's software systems appears to have assigned legally inaccurate expiration dates to over 27,000 patents~\cite{Cheian22}. In 2012, a leap day bug in Microsoft Azure's certificate-handling logic triggered a worldwide outage of the cloud platform~\cite{Laing12}. Twelve years later, in 2024, another leap day bug caused gas pumps in New Zealand to stop dispensing fuel on February~29th~\cite{Reuters24}. As recently as 2026, a bug in timelib, the date library underlying PHP, made equivalent timestamp conversions disagree by one day for dates in February of the year~0~\cite{Gelbmann26}. An empirical study of open-source code~\cite{Tiwari25} confirms that date and time logic is a recurring source of software bugs, which often escape conventional testing due to tricky edge cases. 
Modern artificial intelligence (AI) tools using large language models (LLMs) also suffer from occasional glitches in temporal reasoning~\cite{Fatemi24, Gaere25}, making them untrustworthy for complex document-analysis tasks (e.g., of contracts or legal documents) or validation of calendrical algorithms. Although these issues are well known, there has so far been limited tool support for formally reasoning about date and calendar logic. In fact, recent work applying symbolic execution in the field of computational law~\cite{Goutagny25} explicitly calls out the complexity of date semantics and the lack of tool support to automate such reasoning as a source of its incompleteness.

In other domains with analogous domain-specific problems, such as string-based computations involving regular expressions, the formal methods community has made great strides in the development of string-based constraint solvers~\cite{Kiezun09, Saxena10, Zheng13, Liang14, Trinh14, Berzish17, Zheng17, Mora21} which enable efficient and reliable computer-aided verification. This paper takes inspiration from prior work to introduce support for logical reasoning with common date operations.

We present \datesat{}, a framework for expressing and solving constraint satisfiability problems involving dates and calendar periods. \datesat{} is an extension of quantifier-free linear integer arithmetic that introduces two new sorts \texttt{Date}  and \texttt{Period}, whose elements are constructed as \texttt{Date(year, month, day)} and 
\texttt{Period(years, months, days)} respectively, with support for free date variables, date/period arithmetic operations, and date comparisons. For example, the puzzle above can be expressed with free variables \code{today} and \code{birthdate} as:%
\begin{Verbatim}[fontsize=\footnotesize,commandchars=\\\{\}]
 \PYG{n}{birthdate} \PYG{o}{+} \PYG{n}{Period}\PYG{p}{(}\PYG{l+m+mi}{26}\PYG{p}{,} \PYG{l+m+mi}{0}\PYG{p}{,} \PYG{l+m+mi}{0}\PYG{p}{)} \PYG{o}{\PYGZgt{}} \PYG{n}{today} \PYG{o}{\PYGZhy{}} \PYG{n}{Period}\PYG{p}{(}\PYG{l+m+mi}{0}\PYG{p}{,}\PYG{l+m+mi}{0}\PYG{p}{,}\PYG{l+m+mi}{2}\PYG{p}{)}   \PYG{c+c1}{\PYGZsh{} Not quite 26 two days ago}
\PYG{p}{(}\PYG{n}{birthdate} \PYG{o}{+} \PYG{n}{Period}\PYG{p}{(}\PYG{l+m+mi}{28}\PYG{p}{,} \PYG{l+m+mi}{0}\PYG{p}{,} \PYG{l+m+mi}{0}\PYG{p}{)}\PYG{p}{)}\PYG{o}{.}\PYG{n}{year} \PYG{o}{==} \PYG{n}{today}\PYG{o}{.}\PYG{n}{year} \PYG{o}{+} \PYG{l+m+mi}{1}   \PYG{c+c1}{\PYGZsh{} Will turn 28 next year}
\end{Verbatim}
Here, \texttt{.year} is postfix notation for a selector function that extracts the year component of a \texttt{Date} as an integer.
Spoiler alert: the constraint is \emph{satisfiable} and one solution to the puzzle is the assignment {\small\texttt{today = Date(2026, 1, 1); birthdate = Date(1999, 12, 31)}}. 

This paper formalizes the syntax and semantics of date and period constraints, using the conventions employed by popular programming libraries.
Then, the paper explores various strategies for solving \datesat{} constraints with commodity SMT solvers~\cite{Barrett18}, via reduction to the theory of linear integer arithmetic, which turns out to be non-trivial. A \emph{\naive} approach of representing dates as 3-tuples $\Date{y,m,d}$ leads to complex inefficient integer constraints when handling cases where the day ``overflows'' past the end of a month (e.g., adding ``100~days'' to a symbolic $\Date{y,m,d}$ requires updating not just $d$ but also $m$ and potentially $y$). Representing dates as $\Delta$ days since an \emph{epoch}---analogous to Unix timestamps---makes day-addition easy, but leads to inefficient calculations involving months and years (e.g., adding ``6 months'' to $\Delta$ requires knowing where $\Delta$ is in the calendar). 
We present four encoding strategies that trade off between these costs, including a hybrid one that lazily maintains both representations to make both year/month and day arithmetic efficient.

We implement these strategies in Python using the Z3~\cite{Z3} solver's Python API. To evaluate the strategies, we prepare a dataset of \totalConstraints{} \datesat{} constraints synthesized with LLMs, randomized grammar-based sampling, and mining of legal statutes; we call this dataset \dataset. No single encoding is best on \dataset{}: a variant of the hybrid encoding solves the most constraints within the 20\,s timeout (93.3\%), but the epoch-based encoding is the fastest on the most constraints (350 of the 768 that any encoding solves, over twice as many as any other), and every encoding is the fastest on some. We therefore train a router that picks an encoding for each constraint dynamically based on its features; it solves 94.8\% of the constraints, more than any single encoding, and its geometric-mean speedup over the \naive{} encoding, $5.47\times$, is also the highest.

In summary, the contributions of this work include:
\begin{enumerate}
\item A formalization of the date and period constraint satisfiability problem, based on prior work on formalizing date and period semantics~\cite{Monat24}. (\S\ref{sec:semantics})
\item Four strategies for solving \datesat{} constraints, differing in representation formats and algorithms for efficiently encoding date and period arithmetic as integer constraints. (\S\ref{sec:solver})
\item A mechanized specification of \datesat{} semantics and of each solver encoding in Verus, with machine-checked proofs of equisatisfiability. (\S\ref{sec:correctness}; proofs at {\small\url{\datesatrepo/tree/main/proofs}})
\item An open-source implementation of \datesat{} in Python, using API bindings for the Z3 SMT solver. ({\small\url{\datesatrepo}})
\item A benchmark suite of \totalConstraints{} \datesat{} constraints compiled from synthetic and real-world sources. (\S\ref{sec:dataset}; {\small\url{\datesatbenchrepo}})
\item An empirical evaluation of the \datesat{} solvers on the benchmark suite, showing that the encodings have complementary strengths, and a router that leverages them by dynamically picking an encoding for each constraint. (\S\ref{sec:evaluation})
\end{enumerate}

\noindent We are happy to report that since open-sourcing \datesat{} earlier this year, our proposed encodings are being incorporated\footnote{Hypothesis issue \hypissue{4759} and merged PR \hyppr{4760}; CrossHair issues \chissue{424}, \chissue{428}, \chissue{431} and merged PRs \chpr{425}, \chpr{427}.} into open-source tools like Python's Hypothesis property-based testing library (9.2k~GitHub stars) and the CrossHair symbolic analysis engine (1.4k~GitHub stars) by their respective maintainers, who referenced our work independently.
\section{Motivating Examples}
\label{sec:motivation}

We motivate \datesat with two real-world use cases: software refactoring and legal compliance.

\begin{figure}[t]
\centering
\begin{tcolorbox}[
  enhanced,
  colback=white,
  colframe=black,
  boxrule=0.4pt,
  sharp corners,
  left=1pt,right=1pt,top=2pt,bottom=2pt
]
\begin{Verbatim}[
  fontsize=\scriptsize,
  linenos,
  breaklines=true,
  breakanywhere=true,
  numbersep=6pt,
  commandchars=\\\{\}
]
\PYG{k+kn}{from}\PYG{+w}{ }\PYG{n+nn}{datetime}\PYG{+w}{ }\PYG{k+kn}{import} \PYG{n}{date}
\PYG{k+kn}{from}\PYG{+w}{ }\PYG{n+nn}{dateutil}\PYG{n+nn}{.}\PYG{n+nn}{relativedelta}\PYG{+w}{ }\PYG{k+kn}{import} \PYG{n}{relativedelta}

\PYG{k}{def}\PYG{+w}{ }\PYG{n+nf}{is\PYGZus{}in\PYGZus{}same\PYGZus{}18m\PYGZus{}window\PYGZus{}1}\PYG{p}{(}\PYG{n}{base\PYGZus{}date}\PYG{p}{:} \PYG{n}{date}\PYG{p}{,} \PYG{n}{event\PYGZus{}date}\PYG{p}{:} \PYG{n}{date}\PYG{p}{)} \PYG{o}{\PYGZhy{}}\PYG{o}{\PYGZgt{}} \PYG{n+nb}{bool}\PYG{p}{:}
    \PYG{k}{if} \PYG{n}{event\PYGZus{}date} \PYG{o}{\PYGZlt{}} \PYG{n}{base\PYGZus{}date}\PYG{p}{:}
        \PYG{k}{return} \PYG{k+kc}{False}

    \PYG{n}{elapsed\PYGZus{}m} \PYG{o}{=} \PYG{p}{(}\PYG{n}{event\PYGZus{}date}\PYG{o}{.}\PYG{n}{year} \PYG{o}{\PYGZhy{}} \PYG{n}{base\PYGZus{}date}\PYG{o}{.}\PYG{n}{year}\PYG{p}{)}\PYG{o}{*}\PYG{l+m+mi}{12} \PYG{o}{+} \PYG{p}{(}\PYG{n}{event\PYGZus{}date}\PYG{o}{.}\PYG{n}{month} \PYG{o}{\PYGZhy{}} \PYG{n}{base\PYGZus{}date}\PYG{o}{.}\PYG{n}{month}\PYG{p}{)}
    \PYG{k}{if} \PYG{n}{event\PYGZus{}date}\PYG{o}{.}\PYG{n}{day} \PYG{o}{\PYGZlt{}} \PYG{n}{base\PYGZus{}date}\PYG{o}{.}\PYG{n}{day}\PYG{p}{:}
        \PYG{n}{elapsed\PYGZus{}m} \PYG{o}{\PYGZhy{}}\PYG{o}{=} \PYG{l+m+mi}{1}

    \PYG{k}{return} \PYG{n}{elapsed\PYGZus{}m} \PYG{o}{\PYGZlt{}} \PYG{l+m+mi}{18}

\PYG{k}{def}\PYG{+w}{ }\PYG{n+nf}{is\PYGZus{}in\PYGZus{}same\PYGZus{}18m\PYGZus{}window\PYGZus{}2}\PYG{p}{(}\PYG{n}{base\PYGZus{}date}\PYG{p}{:} \PYG{n}{date}\PYG{p}{,} \PYG{n}{event\PYGZus{}date}\PYG{p}{:} \PYG{n}{date}\PYG{p}{)} \PYG{o}{\PYGZhy{}}\PYG{o}{\PYGZgt{}} \PYG{n+nb}{bool}\PYG{p}{:}
    \PYG{k}{if} \PYG{n}{event\PYGZus{}date} \PYG{o}{\PYGZlt{}} \PYG{n}{base\PYGZus{}date}\PYG{p}{:}
        \PYG{k}{return} \PYG{k+kc}{False}

    \PYG{n}{window\PYGZus{}end} \PYG{o}{=} \PYG{n}{base\PYGZus{}date} \PYG{o}{+} \PYG{n}{relativedelta}\PYG{p}{(}\PYG{n}{months}\PYG{o}{=}\PYG{l+m+mi}{18}\PYG{p}{)}
    \PYG{k}{return} \PYG{n}{event\PYGZus{}date} \PYG{o}{\PYGZlt{}} \PYG{n}{window\PYGZus{}end}
\end{Verbatim}
\end{tcolorbox}
\caption{Two implementations of a Python function for determining if a given event date is within an 18-month window of a base date. Formally checking equivalence via verification or symbolic execution requires reasoning about date and period arithmetic.}
\label{fig:motivating-example-refactoring}
\end{figure}

Fig.~\ref{fig:motivating-example-refactoring} lists two Python functions intended to determine if an \texttt{event\_date} falls within an 18-month window of a starting \texttt{base\_date}. Both of these were written by ChatGPT: the first function attempts this without any external libraries---by computing the whole number of \emph{elapsed months} between the two dates and checking that it is less than 18---whereas the second version uses the \code{dateutil} library to find the end of the 18-month window and checks that the event is within it. Now, are these two functions equivalent? We used the \emph{Hypothesis}~\cite{MacIver2019Hypothesis} property-based testing tool to randomly generate ten thousand random pairs of \code{base\_date} and \code{event\_date} and assert the equivalence of these two functions---no errors were found. However, there actually exists a subtle edge case when \code{base\_date} is say ``March 31, 2025'' and \code{event\_date} is ``September 30, 2026''. When the second function computes \code{window\_end} by adding 18 months, there is no ``September 31, 2026'', and so it is rounded down to ``September 30, 2026'' by default%
, so the second function returns \code{False} while the first returns \code{True}. The two functions are not equivalent, and such semantic nuances easily escape testing. One could potentially hope to find this discrepancy using symbolic execution or solver-aided verification, but naively doing so by tracing Python's internal implementations of \code{date}, \code{relativedelta}, etc. would lead to an explosion of constraints. For instance, CrossHair~\cite{CrossHair}, which traces \code{relativedelta} rather than reasoning about period arithmetic directly, finds no difference within 300\,s on dates in the years 1900--2100. Running the example from Fig.~\ref{fig:motivating-example-refactoring} on representative inputs executes 17,907 lines of Python code, comprising 2,723 function calls, 5,934 conditional branches, 1,263 loop iterations, 1,882 comparison operations and 1,206 arithmetic operations\footnote{With Python~3.13.1, using the original \code{datetime} module (i.e. no C code) for tracing.}. Using \datesat{} (ref. Fig.~\ref{fig:dsl-example-refactoring}), the problem fits in about a dozen comparisons and a counter-example is found in
 \emph{0.023\,s}.

\begin{figure}[t]
\centering
\footnotesize
\fbox{%
\begin{minipage}[t]{0.96\linewidth}
\textbf{U.S. Code: Title 26, \S 338---Certain stock purchases treated as asset acquisitions (\emph{selected excerpts})}
\\
\emph{(g)(1) Except as otherwise provided in regulations, an election under this section shall be made not later than the 15th day of the 9th month beginning after the month in which the acquisition date occurs.}

\emph{(h)(2) The term “acquisition date” means, with respect to any corporation, the first day on which there is a qualified stock purchase with respect to the stock of such corporation.}

\emph{(d)(3) The term “qualified stock purchase” means any transaction or series of transactions in which stock (...) of 1 corporation is acquired by another corporation by purchase during the 12-month acquisition period.}

\emph{(h)(1) The term “12-month acquisition period” means the 12-month period beginning with the date of the first acquisition by purchase of stock included in a qualified stock purchase (...)}

\end{minipage}%
}
\caption{Excerpts of legal statutes from the U.S. Internal Revenue Code (IRC) that deal with corporate taxation. Automating legal reasoning tasks requires validating constraints involving date and period arithmetic.}
\label{fig:motivating-example-legal}
\end{figure}

Fig.~\ref{fig:motivating-example-legal} shows an excerpt of legal statutes from the United States Code that deals with federal taxation. It illustrates the time window available to a corporation acquiring another via a stock purchase in order to treat this purchase as an \emph{asset acquisition}, which has different tax treatment than a capital investment. Now suppose we wanted to know ``\emph{Can a corporation performing an acquisition by purchase make a Section 338(g) election 500 days after its first purchase of the target stock?}''. One way of answering this is through the aid of AI tools, which are increasingly applied to legal document analysis and deductive reasoning tasks~\cite{Yoshioka21-DataAugmentation, Goebel23-COLIEE23, Guha24-LegalBench, Nay24-LLMsTaxAttorneys, Blair-Stanek23-GPT3Statutory, Padhye24-SEAI4Law}, though prior work has also noted limitations of LLMs for calendar and temporal reasoning~\cite{Fatemi24, Gaere25}. In our experiment, Claude Sonnet~5.5 (default medium effort) and GPT-6 Luna (medium effort) answered~``No'', reasoning that 500 days exceeds the roughly 8.5-month deadline from the acquisition date.
However, the correct answer is ``Yes'' because the \emph{first purchase} of the stock (e.g., on ``January 12, 2024'') can occur up to 12 months before the acquisition date (e.g., ``December 21, 2024''), implying a corresponding election deadline
(``September 15, 2025'') such that the election date can be 500 days from the first purchase (e.g., on ``May 26, 2025'').
More capable models such as Opus~5.5, Fable~5.1, GPT-6 Astra, and GPT-6.1 Sol do answer correctly, but at substantially higher latency and cost. 
When the \datesat{} solver is provided to Claude Sonnet~5.5 using the Model Context Protocol (MCP), it synthesizes a constraint similar to Fig.~\ref{fig:dsl-example-legal} and produces a correct response.

\section{Problem Formalization}
\label{sec:semantics}

In principle, \datesat{} provides a quantifier-free logic for comparing dates and performing arithmetic involving dates and calendar periods. 

\subsection{Date and Period Semantics}
\label{subsec:semantics}

We begin by formally defining the domain of values and operations thereof supported by \datesat{}.

\begin{definition}[Date]\label{def:date}
A date is represented as a triple of three integers $\Date{y, m, d}$ where $y$ is the year following astronomical year numbering, $m$ is the ordinal of the month, and $d$ is the ordinal of the day in a month\footnote{For clarity, in the text we will often use a verbose notation such as ``February 29, 2024'' which means the value $\Date{2024, 2, 29}.$}. We say a date is \emph{valid} if it respects the rules of the Gregorian calendar, and let $\Dates$ be the set of all valid dates. More formally, if we let $\nbdays{y, m}$ be the number of days in the month $m$ of year $y$ in the Gregorian calendar\footnote{For example, $\nbdays{2024, 2} = 29$ because February has 29 days in the year 2024.}, then :
$$valid(\Date{y, m, d}) \Leftrightarrow 1 \leq m \leq 12 \wedge 1 \leq d \leq \nbdays{y, m}$$
$$\Dates = \{ v \in \mathbb{Z}^3 \mid valid(v)\}$$
\end{definition}

\begin{definition}[Date Comparison]\label{def:date-comparison} We define a total ordering on dates $\langle \Dates, \leq \rangle$ with a lexicographic comparison:
\[
\Date{y_1, m_1, d_1} \leq \Date{y_2, m_2, d_2} \Leftrightarrow 
\begin{cases}
y_1 < y_2 & \text{or} \\
y_1 = y_2 \wedge m_1 < m_2 & \text{or} \\
y_1 = y_2 \wedge m_1 = m_2 \wedge d_1 \leq d_2
\end{cases}
\]
\end{definition}

\begin{definition}[Period]\label{def:period}
We represent a \emph{period} of $n_y$ years, $n_m$ months, and $n_d$ days as a triple of three integers $\Period{n_y, n_m, n_d}$, and let $\Periods = \mathbb{Z}^3$ be the set of all periods.
\end{definition}

\noindent For example, the period of ``1 year, 2 months, and 15 days'' is represented as $\Period{1, 2, 15}$. The fundamental purpose of a period is to perform arithmetic relative to a base date. For example, adding the above period to a base date of ``January~1, 2000'' should produce ``March~16, 2001''. On their own, periods cannot be compared, neither for ordering nor 
for equality. For example, we cannot establish whether the period $\Period{0, 1, 0}$ is smaller, equal to, or larger than the period $\Period{0, 0, 30}$ without knowing which date the period is relative to, since ``1 month'' can have anywhere from 28 to 31 days. Note that the separation between years and months in a period is purely syntactic: a period of $n_y$ years and $n_m$ months is equivalent to one of $0$ years and $12 n_y + n_m$ months, as reflected in the \textsc{Add-Period} rule of Fig.~\ref{fig:semantics}. The distinction between days and months/years, by contrast, is semantic, as the number of days in a month or year depends on the base date.

Periods can, however, be combined with element-wise addition, so $\Period{1, 2, 15} + \Period{0, 0, 15} = \Period{1, 2, 30}$. Periods can also be scaled, so $2 \times \Period{1,2,15} = \Period{2,4,30}$. In \datesat{}, periods are always concrete: no operation on dates produces a period as result; however, periods can be added to dates, as below.

\begin{figure}[!t]
\footnotesize
\begin{mathpar}
\inferrule*[lab=Add-Months]
  {y' = y + \left\lfloor\frac{m-1+n}{12}\right\rfloor \qquad
   m' = 1 + ((m-1+n) \bmod 12)}
  {\Date{y,m,d} \plusMonths n \rightarrow
   \Date{y',\, m',\, \min(d,\nbdays{y',m'})}}
\and
\inferrule*[lab=Add-Days]
  {1 \le d \le \nbdays{y, m} \qquad 1 \le d + n \le \nbdays{y, m}}
  {\Date{y, m, d} \plusDays n \rightarrow \Date{y, m, d+n}}
\\
\inferrule*[lab=Add-Days-Over]
  {1 \le d \le \nbdays{y, m} \qquad d + n > \nbdays{y, m}}
  {\Date{y, m, d} \plusDays n \rightarrow (\Date{y, m, 1} \plusMonths 1) \plusDays (n - (\nbdays{y, m} - d) - 1)}
\and
\inferrule*[lab=Add-Days-Under1]
  {1 < d \le \nbdays{y,m} \qquad d + n \le 0}
  {\Date{y, m, d} \plusDays n \rightarrow \Date{y, m, 1} \plusDays (d - 1 + n)}
\\
\inferrule*[lab=Add-Days-Under2]
  {1 + n \le 0 \qquad \Date{y, m, 1} \plusMonths (-1) \rightarrow \Date{y', m', 1}}
  {\Date{y, m, 1} \plusDays n \rightarrow \Date{y', m', 1} \plusDays (n + \nbdays{y', m'}) }
\and
\inferrule*[lab=Add-Comp]
  {e \rightarrow e'}
  {e \plusEither n \rightarrow e' \plusEither n} \quad{\text{where } \plusEither \text{ is either } \plusMonths \text{ or } \plusDays}
\\
\inferrule*[lab=Add-Period]
  {n_m' = 12 \times n_y + n_m}
  {\Date{y,m,d} \plusPeriod \Period{n_y, n_m, n_d} \rightarrow (\Date{y,m,d} \plusMonths n_m') \plusDays n_d}
\end{mathpar}
\caption{Semantics for date and period arithmetic in \datesat{}, derived from prior work by \citeauthor{Monat24} under the CC-BY 4.0 license. Compared to that work's parametric rounding semantics, \datesat{} assumes a default rounding-down for ambiguous dates that may result when adding months or years.}
\label{fig:semantics}
\end{figure}

\begin{definition}[Date and Period Addition]\label{def:date-period-add}
A date $\Date{y, m, d}$ can be offset by adding a period $\Period{n_y, n_m, n_d}$ resulting in a new date $\Date{y', m', d'}$. We define the binary operator $\plusPeriod : \Dates \times \Periods \rightarrow \Dates$ for this operation, whose small-step semantics are provided in Fig.~\ref{fig:semantics}. 
The result of date-period addition is the unique valid date reached by applying the small-step rules repeatedly, starting with Add-Period (see Theorem~\ref{thm:normalization}).
\end{definition}

\noindent The crux of precisely defining semantics for date and period arithmetic is in dealing with invalid dates which result from naive element-wise operations. For example, consider $\Date{2017, 12, 30} \plusPeriod \Period{2, 2, 1}$. There are several possible outcomes depending on the order in which elements are added as well as on how invalid dates are handled.
The convention we follow, which is consistent with widely used programming libraries\footnote{E.g., Java's \code{java.time}, Python's \code{dateutil}, JavaScript's \code{Temporal}, Rust's \code{chrono}, and \code{INTERVAL} addition in PostgreSQL and MySQL.}, is to first perform year/month addition with overflows followed by round-down of invalid dates ($\plusMonths$), then perform day addition ($\plusDays$) allowing any overflows to adjacent months.
In the prior example, we would first add ``2 years and 2 months''---i.e., 26 months---to get $\Date{2020, 2, 30}$ (after handling overflows), then round this down to a valid date of $\Date{2020, 2, 29}$, and then add ``1 day'' to get $\Date{2020, 3, 1}$ (after handling overflows). Negative periods, and equivalently date-period subtraction, are handled similarly.

The semantics of Fig.~\ref{fig:semantics} are a simplified version of the semantics formalized in prior work by \citet{Monat24}. While they provide a parametric framework for resolving invalid dates after year/month addition (e.g., rounding up, rounding down, or raising an error), we fix the semantics to always \emph{rounding down} for the reason mentioned above; Fig.~\ref{fig:semantics} is what results when this choice is inlined into their formalization. The prior work also presents several theorems about the formalization, for example that adding any period from a valid date results in a valid date, restated in our notation:

\begin{theorem}[Normalization and Well-Formedness]
\label{thm:normalization}
For any valid date $D \in \Dates$ and any period $P \in \Periods$, there exists a (unique) valid date $D' \in \Dates$ such that $D \plusPeriod P \overset{*}{\rightarrow} D'$.
\end{theorem}
\noindent From now on, we abuse the notation $\plusPeriod$ to mean the normalized result of the date and period addition. We also re-state some other notable properties and non-properties about the arithmetic:

\begin{property}[Monotonicity]
For any two dates $D_1, D_2 \in \Dates$ and any period $P \in \Periods$, we have $D_1 \leq D_2 \Rightarrow (D_1 \plusPeriod P) \leq (D_2 \plusPeriod P)$.
\end{property}

\noindent Note that the monotonicity is not strict because we can have dates $D_1 = \Date{2020, 1, 30}$ and $D_2 = \Date{2020, 1, 31}$ where $D_1 < D_2$ but when adding a period of $\Period{0, 1, 0}$ they both result in $\Date{2020, 2, 29}$.

\begin{property}[Non-commutativity]
It is not the case that for all dates $D \in \Dates$ and periods $P_1, P_2 \in \Periods$ that $(D \plusPeriod P_1) \plusPeriod P_2 = (D \plusPeriod P_2) \plusPeriod P_1$.
\end{property}

\noindent For example, when $D = \Date{2020, 1, 30}$, $P_1 = \Period{0, 1, 0}$, and $P_2 = \Period{0, 0, 1}$, we have $D \plusPeriod P_1 \plusPeriod P_2 = \Date{2020, 3, 1}$ (adding ``1 month'' first rounds to ``February 29'', and then adding ``1 day'' results in ``March 1'') but $D \plusPeriod P_2 \plusPeriod P_1 = \Date{2020, 2, 29}$ (adding ``1 day'' first results in ``January 31'', which when advanced by ``1 month'' is ``February 29'').

\begin{property}[Non-associativity]
It is not the case that for all dates $D \in \Dates$ and periods $P_1, P_2 \in \Periods$ that $(D \plusPeriod P_1) \plusPeriod P2 = D \plusPeriod (P_1 + P_2)$.
\end{property}

\noindent For example, when $D = \Date{2020, 2, 29}$, $P_1 = \Period{1, 0, 0}$, and $P_2 = \Period{0, 1, 0}$, if we first add ``1 year'' we round to ``February 28, 2021'' and then adding ``1 month'' results in ``March 28, 2021''. However, adding the period of ``1 year, 1 month'' directly results in ``March 29, 2021''.

\begin{property}[Non-invertibility]
It is not the case that for all dates $D \in \Dates$ and periods $P \in \Periods$ that $(D \plusPeriod P) \plusPeriod (-1 \times P) = D$.
\end{property}

\noindent For example, when $D = \Date{2020, 1, 31}$ and $P = \Period{0, 1, 0}$, adding ``1 month'' rounds down to ``February 29, 2020'', and subtracting ``1 month'' then results in ``January 29, 2020''.

\subsection{Input Language for Constraint Satisfaction Problem}
\label{subsec:dsl}

\begin{figure}[t]
\begin{subfigure}[t]{0.52\textwidth}
\centering
\begin{Verbatim}[fontsize=\scriptsize,linenos=true,
        frame=lines,
        numbersep=5pt,
        breaklines=true,
        breakanywhere=true,
  commandchars=\\\{\}
]
base, event, window\PYGZus{}end : Date
elapsed\PYGZus{}m, elapsed\PYGZus{}m\PYGZus{}adj: int
result\PYGZus{}1, result\PYGZus{}2: bool
base \PYGZlt{}= event  \PYGZsh{} assume for simplicity

elapsed\PYGZus{}m == (event.year \PYGZhy{} base.year) * 12
      + (event.month \PYGZhy{} base.month)
((event.day \PYGZlt{} base.day)  \PYGZhy{}\PYGZgt{}
      (elapsed\PYGZus{}m\PYGZus{}adj == elapsed\PYGZus{}m \PYGZhy{} 1))
((event.day \PYGZgt{}= base.day) \PYGZhy{}\PYGZgt{}
      (elapsed\PYGZus{}m\PYGZus{}adj == elapsed\PYGZus{}m))
result\PYGZus{}1 == (elapsed\PYGZus{}m\PYGZus{}adj \PYGZlt{} 18)

window\PYGZus{}end == base + Period(0, 18, 0)
result\PYGZus{}2 == (event \PYGZlt{} window\PYGZus{}end)

result\PYGZus{}1 != result\PYGZus{}2  \PYGZsh{} Counter\PYGZhy{}example?
\end{Verbatim}
\caption{A \datesat{} constraint checking for counter-examples to the refactoring shown in Fig.~\ref{fig:motivating-example-refactoring}. If satisfiable, then the two Python functions are not equivalent.}
\label{fig:dsl-example-refactoring}
\end{subfigure}
\hfill
\begin{subfigure}[t]{0.42\textwidth}
\centering
\begin{Verbatim}[fontsize=\scriptsize,linenos=true,
        frame=lines,
        numbersep=5pt,
        breaklines=true,
        breakanywhere=true,
        breakautoindent=true,
        breaksymbolleft={},
        breaksymbolright={},
  commandchars=\\\{\}
]
first\PYGZus{}buy : Date
acq\PYGZus{}date : Date
elec\PYGZus{}ddl : Date
elec\PYGZus{}date : Date

acq\PYGZus{}date \PYGZgt{}= first\PYGZus{}buy
acq\PYGZus{}date \PYGZlt{} first\PYGZus{}buy +
           Period(0, 12, 0)
elec\PYGZus{}ddl.day == 15
elec\PYGZus{}ddl.year == (acq\PYGZus{}date +
           Period(0, 9, 0)).year
elec\PYGZus{}ddl.month == (acq\PYGZus{}date +
           Period(0, 9, 0)).month
elec\PYGZus{}date \PYGZlt{}= elec\PYGZus{}ddl

elec\PYGZus{}date == first\PYGZus{}buy
           + Period(0, 0, 500)
\end{Verbatim}
\caption{A \datesat{} constraint for checking if the rules from Fig.~\ref{fig:motivating-example-legal} allow a \S338(g) election to be made 500 days after the first purchase.} 
\label{fig:dsl-example-legal}
\end{subfigure}

\caption{\datesat{} constraints for the motivating examples. }
\label{fig:dsl-examples}
\end{figure}

In practice, we are interested in solving constraints that may involve a mix of boolean logic, integer arithmetic, and date and period operations. We therefore extend the logic of quantifier-free linear integer arithmetic to include dates, date comparisons, and date/period arithmetic. Fig.~\ref{fig:dsl-examples} demonstrates how the motivating examples (Figs.~\ref{fig:motivating-example-refactoring} and \ref{fig:motivating-example-legal}) can be encoded as constraints in \datesat{}. The full syntax of the constraint language supported by \datesat{} is provided in Fig.~\ref{fig:dsl-grammar}. Succinctly, the language extends quantifier-free boolean logic and linear integer arithmetic (QF-LIA) with:
(1)~variables of the \texttt{Date} sort; (2)~constructors for \texttt{Date} values from integer literals\footnote{\label{fn:parametric}The language can be extended to support date constructors with arbitrary integer expressions without loss of decidability: every expression of the form \code{Date(e1, e2, e3)} can be replaced with an expression \code{x} for fresh date variable \code{x}, after separately adding constraints \code{x.year == e1 \&\& x.month == e2 \&\& x.day == e3}. While our implementation supports the extended syntax, we use the restricted form to formally reason about correctness.
}; (3)~constructors for \texttt{Period} values from integer literals; (4)~relational comparison between dates (Def.~\ref{def:date-comparison}); (5)~date and period addition/subtraction resulting in a new date (Def.~\ref{def:date-period-add}); (6)~adding or subtracting two periods and (7)~scaling periods by a constant integer factor (Def.~\ref{def:period}); and (8)~extracting integer components (\texttt{year}, \texttt{month}, \texttt{day}) from date expressions.

\begin{figure}[t]
\centering
\begin{tcolorbox}[
  enhanced,
  frame empty,
  colback=white,
  sharp corners,
  left=2pt,right=2pt,top=2pt,bottom=2pt,
  borderline south={0.4pt}{0pt}{black}
]
\begin{Verbatim}[
  linenos,
  frame=none,
  numbersep=6pt,
  fontsize=\scriptsize,
  highlightlines={14-24, 32-36},
  highlightcolor=grammarhighlight,
  commandchars=\\\{\}
]
constraint: bool\PYGZus{}expr

bool\PYGZus{}expr: implication
implication: equality | equality \PYGZdq{}\PYGZhy{}\PYGZgt{}\PYGZdq{} implication
equality: disjunction | disjunction \PYGZdq{}==\PYGZdq{} disjunction | disjunction \PYGZdq{}!=\PYGZdq{} disjunction
disjunction: conjunction | disjunction \PYGZdq{}||\PYGZdq{} conjunction
conjunction: negation | conjunction \PYGZdq{}\PYGZam{}\PYGZam{}\PYGZdq{} negation
negation: \PYGZdq{}!\PYGZdq{} negation | comparison | atom
atom: \PYGZdq{}True\PYGZdq{} | \PYGZdq{}False\PYGZdq{} | BOOL\PYGZus{}VAR | \PYGZdq{}(\PYGZdq{} bool\PYGZus{}expr \PYGZdq{})\PYGZdq{}

CMP\PYGZus{}OP: \PYGZdq{}\PYGZlt{}\PYGZdq{} | \PYGZdq{}\PYGZlt{}=\PYGZdq{} | \PYGZdq{}\PYGZgt{}\PYGZdq{} | \PYGZdq{}\PYGZgt{}=\PYGZdq{} | \PYGZdq{}==\PYGZdq{} | \PYGZdq{}!=\PYGZdq{}

comparison: int\PYGZus{}expr CMP\PYGZus{}OP int\PYGZus{}expr
          | date\PYGZus{}expr CMP\PYGZus{}OP date\PYGZus{}expr

DATE: \PYGZdq{}Date\PYGZdq{} \PYGZdq{}(\PYGZdq{} INTEGER \PYGZdq{},\PYGZdq{} INTEGER \PYGZdq{},\PYGZdq{} INTEGER \PYGZdq{})\PYGZdq{}
PERIOD: \PYGZdq{}Period\PYGZdq{} \PYGZdq{}(\PYGZdq{} INTEGER \PYGZdq{},\PYGZdq{} INTEGER \PYGZdq{},\PYGZdq{} INTEGER \PYGZdq{})\PYGZdq{}

period\PYGZus{}expr: period\PYGZus{}expr \PYGZdq{}+\PYGZdq{} period\PYGZus{}expr | period\PYGZus{}expr \PYGZdq{}\PYGZhy{}\PYGZdq{} period\PYGZus{}expr
           | INTEGER \PYGZdq{}*\PYGZdq{} period\PYGZus{}expr | period\PYGZus{}expr \PYGZdq{}*\PYGZdq{} INTEGER
           | PERIOD | \PYGZdq{}(\PYGZdq{} period\PYGZus{}expr \PYGZdq{})\PYGZdq{}

date\PYGZus{}expr: date\PYGZus{}expr \PYGZdq{}+\PYGZdq{} period\PYGZus{}expr | date\PYGZus{}expr \PYGZdq{}\PYGZhy{}\PYGZdq{} period\PYGZus{}expr
         | DATE | DATE\PYGZus{}VAR | \PYGZdq{}(\PYGZdq{} date\PYGZus{}expr \PYGZdq{})\PYGZdq{}

int\PYGZus{}expr: int\PYGZus{}expr \PYGZdq{}+\PYGZdq{} int\PYGZus{}term | int\PYGZus{}expr \PYGZdq{}\PYGZhy{}\PYGZdq{} int\PYGZus{}term | int\PYGZus{}term
int\PYGZus{}term: INTEGER \PYGZdq{}*\PYGZdq{} int\PYGZus{}term | int\PYGZus{}term \PYGZdq{}*\PYGZdq{} INTEGER | int\PYGZus{}factor
int\PYGZus{}factor: \PYGZdq{}\PYGZhy{}\PYGZdq{} int\PYGZus{}factor
           | INTEGER
           | INT\PYGZus{}VAR
           | \PYGZdq{}(\PYGZdq{} int\PYGZus{}expr \PYGZdq{})\PYGZdq{}
           | date\PYGZus{}expr \PYGZdq{}.\PYGZdq{} DATE\PYGZus{}FIELD

DATE\PYGZus{}FIELD: \PYGZdq{}year\PYGZdq{} | \PYGZdq{}month\PYGZdq{} | \PYGZdq{}day\PYGZdq{}

DATE\PYGZus{}VAR: \PYGZob{}identifier\PYGZcb{}
BOOL\PYGZus{}VAR: \PYGZob{}identifier\PYGZcb{}
INT\PYGZus{}VAR:  \PYGZob{}identifier\PYGZcb{}
INTEGER:  \PYGZob{}signed\PYGZus{}number\PYGZcb{}
\end{Verbatim}
\end{tcolorbox}
\caption{\datesat{} constraint language grammar. The base language is an infix syntax for expressing quantifier-free integer linear arithmetic and boolean logic, whereas the highlighted lines add specific terms for expressing dates, periods, date comparisons, date/period arithmetic, and extracting components of dates.}
\label{fig:dsl-grammar}
\end{figure}

\noindent Given a set of constraints in this language, the goal of a \datesat{} solver is to determine whether a \emph{satisfiable} (SAT) solution exists or if the constraints are \emph{unsatisfiable} (UNSAT)
given the semantics of Sec.~\ref{subsec:semantics}, and, if SAT, to give a satisfying assignment to all Boolean, integer, and date variables.%

\subsection{Evaluation and Satisfiability of Constraints}
\label{subsec:sat}

\begin{definition}[Well-formedness of $\datesat$ constraints]
\label{def:well-formed}
We say that a $\datesat$ constraint is well-formed if for every literal expression \code{Date(y,m,d)}, we have that $valid(\Date{y, m, d})$ holds true as per Definition~\ref{def:date}.
\end{definition}

\begin{definition}[Properly closed \datesat{} constraints]
\label{def:properly-closed}
    Let $E$ be an environment that maps identifiers to concrete values (integers, booleans, or dates). We say a well-formed $\datesat$ constraint is properly closed with environment $E$ if (i) for all \code{INT\_VAR} expressions \code{x}, the mapping $E(x)$ exists and results in an integer value, (ii) for all \code{BOOL\_VAR} expressions \code{y}, the mapping $E(y)$ exists and results in a boolean value, and (iii) for all \code{DATE\_VAR} expressions \code{z}, the mapping $E(z)$ exists and results in a \emph{valid} date as per Definition~\ref{def:date}.
\end{definition}

\begin{definition}[Evaluation of \datesat{} constraints]
\label{def:evaluation}
Given an environment $E$, a \datesat{} constraint $C$ is evaluated by replacing every variable $x$ in $C$ with $E(x)$ and repeatedly reducing sub-expressions using the standard semantics of Boolean algebra and linear integer arithmetic and our semantics for date comparison (Definition~\ref{def:date-comparison}) and date-period arithmetic (Definition~\ref{def:date-period-add}).%
\end{definition}

\begin{lemma}[Interpretability]
\label{lem:interpretability}
A well-formed \datesat{} constraint $C$ that is properly closed by environment $E$ evaluates to a unique boolean value \emph{true} or \emph{false} (i.e., does not get stuck).
\begin{proof}
    By structural induction on the syntax of \datesat{} constraints. First, we show that all date expressions can be reduced to valid date values of the form $\Date{y, m, d}$, because (i) all date literals are valid by Definition~\ref{def:well-formed}, (ii) all date variable references resolve to valid values by Definition~\ref{def:properly-closed}, and (iii) inductively, all date-period arithmetic expressions evaluate to valid date values by Theorem~\ref{thm:normalization}. With this result, we can show that all date comparison expressions can be evaluated to boolean values as per Definition~\ref{def:date-comparison} and all year/month/day-field extractors evaluate to integer values. The rest of the expressions in the \datesat{} grammar represent standard linear-integer arithmetic~\cite{Presburger29} and boolean algebra, and so are trivially interpretable.
\end{proof}
\end{lemma}

\begin{definition}[Satisfiability of \datesat{} constraints]
\label{def:satisfiability}
A \datesat{} constraint $C$ is \emph{satisfiable} iff there exists an environment $E$ such that $C$ is properly closed with $E$ and the evaluation of $C$ with $E$ results in \emph{true}. Conversely, a \datesat{} constraint $C$ is \emph{unsatisfiable} iff for all environments $E$ such that $C$ is properly closed with $E$, the evaluation of $C$ with $E$ results in \emph{false}. 
\end{definition}

By Lemma~\ref{lem:interpretability} and Definition~\ref{def:satisfiability}, all well-formed \datesat{} constraints are either satisfiable or unsatisfiable.

\begin{proposition}[Decidability]
The \datesat{} logic is decidable for well-formed constraints.

\begin{proof}
By encoding of \datesat{} constraints into equisatisfiable constraints in the decidable theory of linear integer arithmetic (ref. Section~\ref{sec:solver}; Theorems~\ref{thm:equisat-naive}--\ref{thm:equisat-ab}).
\end{proof}
\end{proposition}
\section{Solving \datesat{} Constraints, Efficiently}
\label{sec:solver}
We solve \datesat{} constraints by direct translation from the theory of dates (denoted as $\mathcal{T}_{Date}$) to corresponding constraints in $\mathcal{T}_\mathit{LIA}$, the theory of (quantifier-free) linear integer arithmetic extended with divisibility by constants\footnote{This allows operators for Euclidean division ($x\mathbin{\mathrm{div}}n$) and non-negative modulo ($x \bmod n$), for constant $n \neq 0$.}, known to be decidable~\cite{Presburger29, Chistikov24}.\footnote{The encoding strategies do not depend on the target theory, and could also be realized over fixed-width bitvectors. Doing so, however, would bound dates and periods, which are unbounded integers in our semantics, and arithmetic between dates and periods could overflow, requiring guards on every sub-expression. We therefore target linear integer arithmetic.}
Under this paradigm, solver design becomes a problem of defining a mapping $\mathcal{M}: \Phi_{\mathcal{T}_\mathit{Date}} \to \Phi_{\mathcal{T}_\mathit{LIA}}$, where $\Phi_{\mathcal{T}}$ is the domain of constraints in theory $\mathcal{T}$. This mapping must be correct, and its resulting formulas must be efficiently solvable by an SMT solver.
Constructing such a mapping is non-trivial, as calendrical logic is inherently irregular due to varying month lengths and the leap year cycle. Encoding the semantics of dates and periods naively leads to deeply nested if-then-else (ITE) blocks---we use the term \emph{logical depth} to quantify the nesting---and a large number of integer operations---whose cardinality we refer to as \emph{arithmetic complexity}---causing a combinatorial explosion in the search space.%

In this section, we present four encoding strategies, including two variants of the \hybrid encoding, that explore the trade-off between logical depth and arithmetic complexity to optimize solving performance. Sec.~\ref{sec:evaluation} empirically evaluates the five concrete encodings.

\paragraph{Running Example:} Consider the constraint: \code{a + Period(1, 2, 15) < b}, where both \code{a} and \code{b} are free variables of the \code{Date} sort.
This constraint exercises the salient features of \datesat{} including symbolic dates, date-period arithmetic, and date comparison. 
Using this constraint, we show how each mapping $\mathcal{M}$ translates these operations into $\mathcal{T}_\mathit{LIA}$.%

\subsection{\naive Encoding}
\label{sec:naive}
The simplest and most intuitive encoding, $\mathcal{M}_\mathit{simple}$, represents a date $a$ exactly as in Def.~\ref{def:date}: as a tuple of three integer variables ($a_y, a_m, a_d$) for the year, month, and day respectively.
To ensure this represents a valid date according to Def.~\ref{def:date}, $\mathcal{M}_\mathit{simple}$ emits integer constraints $1 \leq a_m \leq 12$ and $1 \leq a_d \leq \nbdays{a_y, a_m}$, where $\nbdays{a_y, a_m}$ is short-hand for $\mathit{ITE}(a_m \in \{4, 6, 9, 11\}, 30, \mathit{ITE}(a_m = 2, \mathit{ITE}(\mathrm{leap}(a_y), 29, 28), 31))$ with $\mathit{leap}(a_y) \Leftrightarrow (a_y \bmod 4 = 0 \land (a_y \bmod 100 \neq 0 \lor a_y \bmod 400 = 0))$.

\paragraph{Date-Period Arithmetic.} 

For our running example of \code{a + Period(1, 2, 15})---or equivalently $a + \Period{1, 2, 15}$ in the formal semantics---we encode a symbolic result date $a'$ via a sequence of intermediate symbolic states: 
\begin{enumerate}
    \item \textbf{Year/Month Addition.} Following the \textsc{Add-Period} rule, we first convert the period's year and month to total months ($12 \times 1 + 2 = 14$) and then apply the \textsc{Add-Months} rule, resulting in a date represented by integer variables $(y_0, m_0, d_0)$. We emit constraints:
    $y_0 = a_y + ((a_m - 1 + 14) \mathbin{\mathrm{div}} 12)$ and $m_0 = ((a_m - 1 + 14) \bmod 12) + 1$. The day is rounded down as needed: $d_0 = \text{ITE}(a_d > \nbdays{y_0, m_0}, \nbdays{y_0, m_0}, a_d)$, with $\nbdays{}$ as before.%
    \item \textbf{Day Addition.} Now to add the 15 days from the period (which is a positive amount) to the symbolic date $(y_0, m_0, d_0)$, we do not know if the result will be in the same month (following the \textsc{Add-Days} rule) or overflow into subsequent months (requiring the \textsc{Add-Days-Over} rule, potentially more than once). We encode this by iteratively computing a sequence of symbolic states $S_0, \dots, S_{15}$, where each $S_i = \Date{y_i, m_i, d_i}$ using fresh integer variables (state $S_0$ was computed above). For each step $i \in [0, 14]$, the state $S_{i+1}$ is determined by first checking if incrementing the day-of-month would cross a month boundary, and if so setting the date to be the 1st of the following month; the latter also requires checking if the month-increment also overflows to January of the following year. Formally, $S_{i+1} = ITE(d_i + 1 > \nbdays{y_i, m_i}, ITE(m_i = 12, \Date{y_i + 1, 1, 1}, \Date{y_i, m_i + 1, 1}), \Date{y_i, m_i, d_i + 1})$, with $\nbdays{}$ expanded as before.  The final result  $a'$ of date and period addition is the last such state, which is $S_{15}$ in our example. When the number of days is negative, a similar encoding of the \textsc{Add-Days-Under-*} rules is performed.
\end{enumerate}

As illustrated, to properly handle overflows, the simple encoding of date/period arithmetic requires a sequence of nested ITE operations proportional to the absolute number of days in the period offset $\lvert n_d \rvert$. The logical depth of the resulting formula is therefore $O(\lvert n_d \rvert)$.\footnote{The \textsc{Add-Days-Over} rule in Fig.~\ref{fig:semantics} performs addition in larger chunks per recursive step of the transition rules, but the recursion depth has the same complexity.} Such formulas pose challenges for SMT solvers, making this encoding inefficient for date/period arithmetic.

\paragraph{Date Comparison.}
Comparisons between date values (e.g., $a' < b$) are mapped to a lexicographical ordering over the integer components, as per Definition~\ref{def:date-comparison}, e.g. $(a'_y < b_y) \lor (a'_y = b_y \land a'_m < b_m) \lor (a'_y = b_y \land a'_m = b_m \land a'_d < b_d)$.

\subsection{\epoch Encoding}
\label{sec:epoch}
The \epoch encoding ($\mathcal{M}_{\textit{epoch}}$) encodes a date value $a$ as a single integer variable $a_{\Delta}$, representing the number of days elapsed since a fixed epoch $\eta$, similar to the representation of Unix time. The motivation for this is to simplify day-only addition and date comparisons to single integer operations. But, what epoch should we choose? We would like to reduce the complexity of the integer arithmetic involved in encoding more complex date/period operations. 
To that end, we choose the epoch date $\eta$ as ``March 1, 2000''. Starting each year in March places the leap day (``Feb 29'') at the end of the year, and $\eta$ starts a 400-year Gregorian cycle of 146,097 days, within which leap days recur every 1461 days ($365 \times 4 + 1$) except in century years not divisible by 400.

\paragraph{Constructing a date value}

Given a \naive date value of the form $\Date{a_y, a_m, a_d}$, we can convert to the corresponding offset $a_\Delta$ as follows: (i) we first compute the number of years elapsed since the epoch (taking into account that years start in March) as $ye$ and the months elapsed in the given year (treating March=0 to February=11) as $me$ like so:
\begin{equation}
  \begin{aligned}
    ye &= \begin{cases} 
      a_y - 2000, & \text{if } a_m \geq 3 \\
      a_y - 2001,   & \text{otherwise}
    \end{cases}
  \end{aligned}
  \qquad
  \begin{aligned}
    me &= \begin{cases} 
      a_m - 3, & \text{if } a_m \geq 3 \\
      a_m + 9, & \text{otherwise}
    \end{cases}
  \end{aligned}
\end{equation}
\noindent Eventually, we want a representation that is only a number of days since the epoch. So, (ii) we compute the number of days attributed to years-elapsed-since-epoch, which is $yd = 365 \times ye + (ye/4 - ye/100 + ye/400)$, the latter being a leap-year correction using Euclidean division; (iii) we compute the number of days attributed to the month component, which is $md = (153 \times me + 2)/5$---
this closed form holds because years start in March;\footnote{Starting in March, the number of days per month is [31, 30, 31, 30, 31, 31, 30, 31, 30, 31, 31, 28 or 29], so the cumulative days elapsed per month is [0, 31, 61, 92, 122, 153, 184, 214, 245, 275, 306, 337], which the expression computes for $me \in [0, 11]$.} and (iv) finally compute $a_\Delta = yd + md + (a_d - 1)$.

\paragraph{Date-Period Arithmetic.}
How $\mathcal{M}_{\textit{epoch}}$ handles date and period arithmetic depends on the components of the period $\Period{n_y, n_m, n_d}$:
\begin{enumerate}
    \item \textbf{Only Adding Days.} If a period contains only a day component $n_d$, the arithmetic is a trivial integer addition: $a'_{\Delta} = a_{\Delta} + n_d$. This reduces the $O(\lvert n_d \rvert)$ logical depth as in the \naive encoding to $O(1)$ arithmetic complexity.
    \item \textbf{Arithmetic Involving Year/Month.} 
    If $n_y$ or $n_m$ are non-zero (e.g., adding ``6 months''), we cannot adjust $a_\Delta$ directly without knowing which calendar year and month it represents (e.g., to determine how many days ``6 months'' means). So, we must temporarily recover the corresponding triple representation $\Date{a_y, a_m, a_d} = \eta \plusPeriod \Period{0, 0, a_{\Delta}}$--- we split $a_\Delta$ into a 400-year cycle $a_\Delta \mathbin{\mathrm{div}} 146097$ and a day of cycle $c = a_\Delta \bmod 146097$. From $c$ we peel off the century, the 4-year block within it, and the year within that block by Euclidean division by 36524, 1461, and 365 in turn, clamping the century and year quotients to at most 3 so that the last day of the cycle and the leap day closing a 4-year block are counted in their own block. The year of cycle is then $100 \times \mathit{century} + 4 \times \mathit{block} + \mathit{year}$, and the month and day are recovered from the remaining day of year by inverting the month expression above. Once the components are recovered, $n_y \times 12 + n_m$ months are added using the same algorithm as the \naive approach, which require a constant number of operations. The resulting triple is then re-encoded into an integer $a''_{\Delta}$ representing days since epoch $\eta$ via the conversion described above. Finally, $n_d$ is added via simple integer addition if needed to get the final result, so $a'_{\Delta} = n_d + a''_{\Delta}$.
\end{enumerate}

\paragraph{Date Comparison.} In this encoding, a date comparison $a' < b$ is reduced to a single integer comparison $a'_{\Delta} < b_{\Delta}$ in $\mathcal{T}_\mathit{LIA}$, eliminating the disjunctions required by lexicographical ordering as in the \naive encoding.

The main advantage of the \epoch encoding is that the addition of the days component does not require deeply nested if-then-else clauses for determining overflows across month boundaries, as in the \naive encoding. However, the trade-off is that arithmetic with months and years requires translation to and from the \naive encoding, shifting the solver's burden from managing logical depth (ITE branching) to handling more integer division and modulo constraints.

\newcommand{\ymdFlag}[1]{#1_\mathit{ymd}}
\newcommand{\deltaFlag}[1]{#1_\eta}
\newcommand{\ymdFix}{\mathrm{fix}_\mathit{ymd}}
\newcommand{\deltaFix}{\mathrm{fix}_\mathit{\eta}}

\subsection{\hybrid Encoding}
\label{sec:hybrid}
The \hybrid encoding ($\mathcal{M}_{\textit{hybrid}}$) attempts to minimize both logical depth and arithmetic complexity by supporting two forms of representing a date: the \naive encoding's triple
$(a_y, a_m, a_d)$ and the \epoch encoding's offset
$a_\Delta$. The key idea is to use the \naive encoding's triple for efficient year/month arithmetic, and the \epoch offset for efficient day arithmetic and comparisons.

\paragraph{Constructing a date value and converting between representations.}
To realize this, the encoder represents each symbolic date value $a$ as a 6-tuple $(a_y, a_m, a_d, a_\Delta, \ymdFlag{a}, \deltaFlag{a})$, where $\ymdFlag{a}$ and $\deltaFlag{a}$ are Python-level Boolean flags tracking the availability or validity of each representation, rather than SMT variables. The flags satisfy the invariant $\ymdFlag{a} \vee \deltaFlag{a} = \true$, with $\ymdFlag{a}$ indicating that $a$ is represented as $\Date{a_y, a_m, a_d}$ and $\deltaFlag{a}$ indicating that $a$ is $(\eta \plusPeriod \Period{0, 0, a_\Delta})$.
The question, then, is which representation to initialize first when constructing a new free date variable. We explored two options: \hybridymd{} initializes $a$ as the triple $(a_y, a_m, a_d)$ as in the \naive encoding (Sec.~\ref{sec:naive}) with the flags $(\ymdFlag{a}, \deltaFlag{a}) = (\true, \false)$, and \hybridepoch{} initializes only the offset $a_\Delta$, as in the \epoch encoding (Sec.~\ref{sec:epoch}), with its flags start as $(\false, \true)$.

After initializing a date variable only with one representation, rather than eagerly keeping both representations consistent on every operation, \hybrid uses a lazy mechanism to keep the representations consistent only as-and-when required. This avoids introducing unnecessary arithmetic into the formula when a sequence of operations can be handled most efficiently entirely within one representation.
During constraint generation, whenever $\ymdFlag{a}$ or $\deltaFlag{a}$ is $\false$ and the corresponding representation is needed for an operation, we ``fix'' the inconsistent or unavailable representation by either converting $a_\Delta$ to $\Date{a_y, a_m, a_d}$ or vice versa using the same logic as described in Sec.~\ref{sec:epoch}, and setting the corresponding flag to $\true$.

\paragraph{Date-Period Arithmetic.}
$\mathcal{M}_{\textit{hybrid}}$ dynamically selects and uses the most efficient representation of a date for the arithmetic based on the period $\Period{n_y, n_m, n_d}$:
\begin{enumerate}
    \item \textbf{Only Adding Days.} If the period contains only a non-zero day component, we use the epoch-based encoding of $a$---this representation is first fixed if $\deltaFlag{a}$ is $\false$ so that $a_{\Delta}$ becomes consistent. Then, $n_d$ is directly added to $a_{\Delta}$, resulting in $a' = (a_y, a_m, a_d, a'_\Delta, \false, \true)$. Note that the $\Date{y,m,d}$ components for such a new intermediate date are not materialized. 
    \item \textbf{Arithmetic Involving Year/Month.} If $n_m$ or $n_y$ is non-zero, we need the
correct component values $\Date{a_y, a_m, a_d}$, which are derived if $\ymdFlag{a} = \false$. Then, $n_y \times 12 + n_m$ months are added using the same method as in Sec.~\ref{sec:naive}. If $n_d = 0$, we are done
and the final result will have flag $\ymdFlag{a'}$ set but flag $\deltaFlag{a'}$ will be $\false$, indicating that the \epoch encoding is inconsistent or not materialized. Otherwise, we follow the procedure in step 1 to add $n_d$ days.
\end{enumerate}

\paragraph{Date Comparison.} 
Date comparisons like $a' < b$ are encoded depending on which representations of $a'$ and $b$ are consistent. If $\deltaFlag{a'} \land \deltaFlag{b}$, then $a'_\Delta < b_\Delta$ suffices. If $\ymdFlag{a'} \land \ymdFlag{b}$, then we emit the lexicographic comparison as in Def.~\ref{def:date-comparison} / Sec.~\ref{sec:naive}. Otherwise, we fix whichever operand ($a'$ or $b$) has an inconsistent \epoch encoding and perform the integer comparison.

\subsection{\ab encoding}
\label{sec:alpha-beta} 
From the previous sections, we observe that year/month arithmetic is easy to encode when we have some information about where in the calendar a date is in, and day arithmetic is easy to encode in an epoch-based representation. The \ab encoding takes a different spin on a hybrid encoding: keeping track of \emph{both} of these views in one concise representation. $\mathcal{M}_{\alpha\beta}$ represents a date as a 2-tuple of integer variables $(a_{\alpha\Delta}, a_{\beta\Delta})$, where $a_{\alpha\Delta}$ represents the number of total months since the epoch $\eta$, and $a_{\beta\Delta}$ represents the number of days elapsed since the start of the month, such that $a_{\beta\Delta} = 0$ corresponds to the first day of a month. In other words, the pair $(a_{\alpha\Delta}, a_{\beta\Delta})$ represents the $(a_{\beta\Delta}+1)$-th day of the month obtained by advancing $a_{\alpha\Delta}$ months from $\eta$. For example, given that our chosen epoch is ``March 1, 2000', the encoding $(2, 5)$ represents ``May 6, 2000''.

\paragraph{Constructing a date value}

Given a \naive date of the form $\Date{a_y, a_m, a_d}$, converting to the corresponding $(a_{\alpha\Delta}, a_{\beta\Delta})$ representation uses basically the same tricks as in the \epoch encoding (Sec.~\ref{sec:epoch}) to compute years-elapsed-since-epoch ($ye$) and months-elapsed-in-the-March-based-year ($me$); then, $a_{\alpha\Delta} = ye \times 12 + me$ and $a_{\beta\Delta} = a_d - 1$.

\paragraph{Date-Period Arithmetic.} Consider the addition $a' = a \plusPeriod \Period{n_y, n_m, n_d}$. Since the integer $a_{\alpha\Delta}$ indicates exactly which year/month a date belongs to, we define $\nbdays{a_{\alpha\Delta}}$ as a short-hand similar to Sec.~\ref{sec:naive}, using ITEs and modulos.
\begin{enumerate}
    \item \textbf{Year/Month Addition.} Following the \textsc{Add-Period} rule, we first compute an intermediate date value $a''$, for which $a''_{\alpha\Delta} = a_{\alpha\Delta} + (12 \times n_y + n_m)$. To ensure that the day remains valid in the new month $a''_{\alpha\Delta}$, we round-down as needed, so $a''_{\beta\Delta} = \min(a_{\beta\Delta}, \nbdays{a''_{\alpha\Delta}} - 1)$.
    \item \textbf{Day Addition.} Now, to add a day component $n_d$ to $a''$, we first compute another intermediate (though potentially invalid) date value $a''' = (a''_{\alpha\Delta}, a''_{\beta\Delta} + n_d$). We check if the second component ($a'''_{\beta\Delta}$) is valid---i.e., in the range $[0,\nbdays{a'''_{\alpha\Delta}})$---and if so we are done with the final result $a' = a'''$. If not, then this invalid $a'''$ is converted to a solely \epoch encoding (as in Sec.~\ref{sec:epoch}) to get a single integer $a'''_\Delta$, which is then re-converted to the \ab encoding to get the final result $a'$. Both these steps can be done by emitting $O(1)$ division and modulo operations, thus avoiding the iterative process of the \naive encoding.
\end{enumerate}

\paragraph{Date Comparison.} Comparisons $a' < b$ are performed via a lexicographical ordering over the pair: $(a'_{\alpha\Delta} < b_{\alpha\Delta}) \lor (a'_{\alpha\Delta} = b_{\alpha\Delta} \land a'_{\beta\Delta} < b_{\beta\Delta})$

\subsection{Correctness of Encodings}
\label{sec:correctness}
We formally verify the correctness of each encoding using Verus~\cite{Lattuada23-Verus}. 
We mechanize the semantics of Fig.~\ref{fig:semantics} directly\footnote{To simplify the mechanization and proofs, we model a de-sugared subset of the \datesat{} grammar that only considers the relational operators `\code{<}' and `\code{=}', since other comparisons can be re-written in terms of these.}, and show that each encoding---\naive, \epoch, both \hybrid, and \ab---is equisatisfiable with the semantics.

\begin{definition}[Satisfiability with Encoding]
    Let $\varepsilon$ be the name of an encoding strategy---such as ``\naive'' or ``\ab''. Then, for any well-formed $\datesat$ constraint $C$ we say that ``$C$ is SAT by $\varepsilon$'' iff the $\varepsilon$-encoded formula of $C$ (in $\mathcal{T}_\mathit{LIA}$) is satisfiable.
\end{definition}

\begin{theorem}[Equisatisfiability of \naive encoding]
\label{thm:equisat-naive}
   Any well-formed $\datesat$ constraint $C$ is SAT by the \naive encoding iff $C$ is satisfiable.
\end{theorem}

\begin{theorem}[Equisatisfiability of \epoch encoding]
\label{thm:equisat-epoch}
   Any well-formed $\datesat$ constraint $C$ is SAT by the \epoch encoding iff $C$ is satisfiable.
\end{theorem}

\begin{theorem}[Equisatisfiability of \hybridymd encoding]
\label{thm:equisat-hybrid-ymd}
   Any well-formed $\datesat$ constraint $C$ is SAT by the \hybridymd encoding iff $C$ is satisfiable.
\end{theorem}

\begin{theorem}[Equisatisfiability of \hybridepoch encoding]
\label{thm:equisat-hybrid-epoch}
   Any well-formed $\datesat$ constraint $C$ is SAT by the \hybridepoch encoding iff $C$ is satisfiable.
\end{theorem}

\begin{theorem}[Equisatisfiability of \ab encoding]
\label{thm:equisat-ab}
   Any well-formed $\datesat$ constraint $C$ is SAT by the \ab encoding iff $C$ is satisfiable.
\end{theorem}

\begin{proof}[Sketch of Proofs] We establish equisatisfiability by showing that, for any well-formed \datesat{} constraint $C$ and any environment $E$ in which it is properly closed, evaluation under the semantics agrees with evaluation under the target encoding $\varepsilon$. This equivalence is established via structural induction over expressions, propagating a key notion of \emph{congruence}: a relation between a date value and an $\varepsilon$-encoding value asserting that they represent the \emph{same calendar date} (i.e., agree on year, month, and day). As a base case, we show that each encoding's date-literal constructor (i.e., from integer values for year, month, and day) produces a value congruent to the corresponding date constructor. The inductive step shows that date-period addition preserves congruence---that is, given a pair of congruent date values across two encodings, adding the same period yields corresponding congruent results across the encodings. This latter step relies on Theorem~\ref{thm:normalization} and the encoding-specific semantics of date-period arithmetic. The key property of congruence is that it \emph{preserves date comparisons}: if two pairs of values are each congruent, then ordering and equality agree across encodings. Since date comparisons are the critical point where the encoding's internal representation could affect boolean outcomes, this is a key bridge. The remaining cases are straightforward: integer expressions involving year/month/day projections reduce to the fact that congruent values agree on these components, arithmetic combinators follow by induction, and boolean connectives simply propagate equivalence from their sub-expressions. Equisatisfiability follows directly from equivalence of evaluation and the definition of satisfiability (Definition~\ref{def:satisfiability}). 
\end{proof}

\noindent \textbf{Note: } The machine-checkable proofs are at {\small\url{\datesatrepo/tree/main/proofs}}.

\section{\dataset: A Benchmark of Constraints Involving Dates and Calendar Periods}
\label{sec:dataset}

We construct \dataset, a comprehensive dataset of \datesat{} constraints, for evaluating our encodings and establishing a reusable benchmark for future work.
\dataset{} comprises \emph{\totalConstraints{} constraints} derived from three sources, each designed for a different evaluation objective.
By design, the grammar-sampled benchmark is restricted to the core date and period grammar to isolate the encodings' handling of calendar operations. The LLM-synthesized and legally grounded benchmarks exercise additional language features to explore semantic corner cases and express legal requirements, respectively.
All \dataset{} constraints bound each declared date variable to the range \Date{1,1,1} to \Date{9999,12,31}.

\paragraph{LLM-Synthesized Constraints.}
To test the encodings on edge cases, we first construct a benchmark of small constraints that exercise semantic corner cases in calendar arithmetic.
We use an LLM (Claude Sonnet~4.5) to generate synthetic \datesat{} constraints, prompting it with a structured system prompt that specifies the \datesat{} constraint language, the set of supported operations, the generation objective, a rigid output schema, and a small set of manually crafted example constraints for few-shot guidance. Each generation objective in the prompt directs the LLM to emphasize one of five scenarios: (1) year vs.\ day contrasts (adding \code{Period(1, 0, 0)} vs.\ \code{Period(0, 0, 365)} or \code{Period(0, 0, 366)}); (2) month vs.\ day contrasts (adding \code{Period(0, 1, 0)} vs.\ \code{Period(0, 0, 31)}); (3) symbolic variables in date constructors like \code{Date(x, 1, 2)} or \code{Date(x + 1, 2, 1)}; (4) field reads on date variables like \code{x.year}, \code{x.month}, and \code{x.day}; or (5) heavy use of logical operators like \code{\&\&}, \code{||}, and \code{!}.
Generated constraints are checked against the \datesat{} parser in a feedback loop that returns syntactic errors to the LLM for correction.
This process yields \emph{100 benchmark instances}.

\paragraph{Grammar-Sampled Constraints.}
To compare the performance of \datesat{} encodings, we systematically sample constraints with varied size and structure.
We use Fandango~\cite{fandango25}, a grammar-based fuzzer, on the date fragment of the \datesat{} grammar (Fig.~\ref{fig:dsl-grammar}): date variables, date literals, and periods added to or subtracted from dates, combined by comparisons, conjunction, and disjunction. It omits integer and Boolean variables and arithmetic, which every encoding translates identically into Z3's integer and Boolean terms, so they would not distinguish the encodings. To diversify the generated constraints, we adopt the swarm testing~\cite{SwarmTesting} approach: before sampling a batch of constraints, we randomly draw a setting for every generation knob, and that setting holds for the whole batch. These knobs (i) alter the size of the constraints (e.g., the number of conjuncts), (ii) approximately control the solution search space (e.g., the number of date variables, and the fraction of them pinned to a literal date), (iii) specify period properties (e.g., the magnitude of each period field, whether a period can mix days, months, and years, or how many period offsets can be chained onto a single date variable), (iv) control date literals (e.g., how far apart the literals of a constraint can lie, and how likely they are to fall near the end of a month), and (v) control the structure of the constraints themselves (e.g., whether they may contain disjunctions). Using this process, we create a benchmark of \emph{500 unique \datesat{} constraints}, drawn from 100 random knob settings.

\paragraph{Legally Grounded Constraints.}
To demonstrate the practical utility of \datesat{} in real-world settings, we additionally construct a suite of constraints derived from the U.S. Internal Revenue Code (IRC)~\cite{irc-section121}. 
We first parse all IRC sections into structured records and filter them to identify clauses likely to contain calendrical requirements. We filter out clauses mentioning time-of-day or time zones. We then use the same LLM to translate selected statutes and provisions into \datesat{} constraints. Generated constraints are checked against the \datesat{} parser using the same feedback loop for syntactic errors as in the LLM-synthesized benchmark. Integer and boolean variables and logical connectives let these constraints combine calendar arithmetic with numeric limits, conditions under which a rule applies, and exceptions to that rule. The authors manually reviewed the generated constraints and judged them to be plausible examples of calendrical reasoning in real-world applications. Our objective is to evaluate the encodings on such reasoning tasks, rather than to produce complete, legally authoritative formalizations of the source statutes. This benchmark consists of \emph{200 unique \datesat{} constraints}, each corresponding to a real-world example.

\begin{figure}[t]
  \centering
  \begin{subfigure}[b]{0.49\linewidth}
\begin{Verbatim}[fontsize=\scriptsize,linenos=true,
        frame=lines,
        numbersep=5pt,
        breaklines=true,
        breakanywhere=true,
  commandchars=\\\{\},codes={\catcode`\$=3\catcode`\^=7\catcode`\_=8\relax}
]
\PYG{esc}{\dsbhl{dsbA}{base, event, window\_end}} : Date
\PYG{esc}{\dsbhl{dsbA}{elapsed\_m, elapsed\_m\_adj}}: int
\PYG{esc}{\dsbhl{dsbA}{result\_1, result\_2}}: bool
base \PYGZlt{}= event  \PYGZsh{} assume for simplicity

elapsed\PYGZus{}m == (event\PYG{esc}{\dsbhl{dsbD}{.year}} \PYGZhy{} base\PYG{esc}{\dsbhl{dsbD}{.year}}) * 12
      + (event\PYG{esc}{\dsbhl{dsbD}{.month}} \PYGZhy{} base\PYG{esc}{\dsbhl{dsbD}{.month}})
((event\PYG{esc}{\dsbhl{dsbD}{.day}} \PYGZlt{} base\PYG{esc}{\dsbhl{dsbD}{.day}})  \PYG{esc}{\dsbhl{dsbB}{->}}
      (elapsed\PYGZus{}m\PYGZus{}adj == elapsed\PYGZus{}m \PYGZhy{} 1))
((event\PYG{esc}{\dsbhl{dsbD}{.day}} \PYGZgt{}= base\PYG{esc}{\dsbhl{dsbD}{.day}}) \PYG{esc}{\dsbhl{dsbB}{->}}
      (elapsed\PYGZus{}m\PYGZus{}adj == elapsed\PYGZus{}m))
result\PYGZus{}1 == (elapsed\PYGZus{}m\PYGZus{}adj \PYGZlt{} 18)

window\PYGZus{}end == base + \PYG{esc}{\dsbhl{dsbC}{Period(0, 18, 0)}}
result\PYGZus{}2 == (event \PYGZlt{} window\PYGZus{}end)

result\PYGZus{}1 != result\PYGZus{}2  \PYGZsh{} Counter\PYGZhy{}example?
\end{Verbatim}
    {\scriptsize\setlength\tabcolsep{3pt}
    \begin{tabularx}{\linewidth}{@{}l>{\raggedright\arraybackslash}Xlr@{}}
      \toprule
       & \textbf{Feature} & \textbf{Category} & \textbf{Value} \\
      \midrule
      \dsbsw{dsbA} & Free variables (date\,/\,int\,/\,bool) & Search space & 3\,/\,2\,/\,2 \\
      \dsbsw{dsbB} & Case splits (\texttt{->}, \texttt{||}) & Boolean skeleton & 2 \\
      \dsbsw{dsbC} & Largest absolute month field & Date arithmetic & 18 \\
      \dsbsw{dsbD} & Field reads (\texttt{.year}\,/\,\texttt{.month}\,/\,\texttt{.day}) & Component extraction & 2\,/\,2\,/\,4 \\
      \midrule
      \multicolumn{4}{@{}l}{\emph{\ldots{} and 53 more features, in 6 categories}} \\
      \bottomrule
    \end{tabularx}}
    \caption{The constraint from Fig.~\ref{fig:dsl-example-refactoring} and a few of its extracted features. Every feature is read directly off the constraint's text without solving it.}
    \label{fig:features-example}
  \end{subfigure}\hfill
  \begin{subfigure}[b]{0.49\linewidth}
    \includegraphics[width=\linewidth]{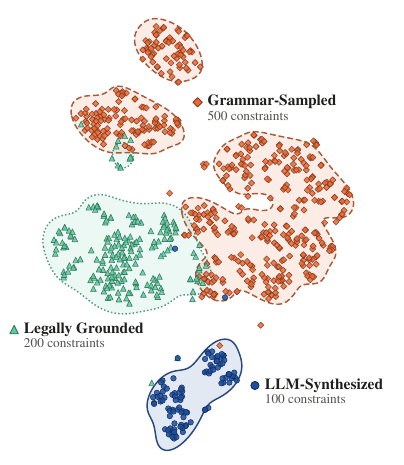}
    \caption{t-SNE projection of all \totalConstraints{} constraints using the 61 nonconstant features. The projection illustrates differences in the structural feature profiles of the three benchmarks.}
    \label{fig:benchmark-tsne}
  \end{subfigure}
  \caption{Characterizing the constraints in \dataset{}.}
  \label{fig:features}
\end{figure}

\paragraph{Characterizing \dataset{}.}
Across all \totalConstraints{} constraints, sizes range from 1 to 32 declared date variables (mean=4.6) and from 3 to 100 comparisons or Boolean variables or literals (mean=19.5).
The benchmarks also exercise different parts of the language: 75\% of the legally grounded constraints declare int or boolean variables, and 68\% contain implications; most LLM-synthesized constraints extract date fields (74\%); and 41\% of the grammar-sampled constraints contain disjunctions.

To characterize the constraints more precisely, we statically extract 61 features from every constraint (listed in full in the \textsc{DateSATBench} repository\footnote{\url{https://github.com/cmu-pasta/DateSATBench/blob/main/analysis/features/features.md}}). The features fall into six categories, each capturing a different aspect of the constraint:
(1)~\emph{search space}: indicators of the size of the space the solver must search and of how its variables are linked, e.g., the number of declared date variables not fixed by unconditional equalities and the number of independent groups of variables, i.e., groups that share no comparison with one another;
(2)~\emph{Boolean skeleton}: the logical shape of the constraint, e.g., the number of date comparisons and the number of case splits (\texttt{||} and \texttt{->}) that the solver may branch on;
(3)~\emph{date arithmetic}: the calendar arithmetic applied to symbolic dates and how it relates them, e.g., the number of date--period operations, the largest absolute month-field value in their periods, and counts of operations comparing a date with itself or with another variable;
(4)~\emph{component extraction}: how often a date is taken apart into its year, month, and day, or built from them, e.g., the number of \code{.day} reads and the number of date constructors with nonliteral components;
(5)~\emph{calendar corners}: indicators of calendar boundary cases, e.g., the presence of February~29 and the fraction of date literals whose day is 28 or later; and
(6)~\emph{encoding costs}: static proxies for the work each encoding performs, e.g., the sum of absolute day offsets applied to symbolic dates for \naive{}, and the number of distinct symbolic date expressions requiring year, month, and day components for \epoch{}.

Fig.~\ref{fig:features-example} shows a few features for the constraint of Fig.~\ref{fig:dsl-example-refactoring}. Fig.~\ref{fig:benchmark-tsne} projects all constraints onto two dimensions with t-SNE~\cite{vandermaaten2008tsne}. We log-transform skewed nonnegative features and standardize all 61 before projection. The three benchmarks occupy largely separate regions.

\section{Evaluation}
\label{sec:evaluation}

To evaluate \datesat{}, we implement all the encoding strategies in Python using the Z3 solver's Python API and run them on \dataset{}. 
We ran all experiments on an isolated Linux server with one Intel Xeon Gold 6226R CPU (16 cores, 2.90\,GHz base clock) and 192\,GB of RAM, running Ubuntu 24.04 LTS (x86\_64), using Python 3.12 and Z3 4.16.0. 
We run Z3 with its default parameters and random seed, one instance at a time in a single process. Each instance has a wall-clock limit of 20\,s, and measured times cover building the encoding, solving, and extracting the satisfying assignment. To account for timing variance, we repeat each experiment 10 times with the same configuration.
Across the 10 runs, each encoding's mean time varies by at most 0.5\%, and each solves the same number of instances in every run.

\begin{table}[t]
\footnotesize
\centering
\caption{Solver outcomes of every encoding on each benchmark of \dataset{}.}
\label{tab:solver-outcomes}
\definecolor{dsSat}{HTML}{2F9E55}
\definecolor{dsUnsat}{HTML}{EFB118}
\definecolor{dsTimeout}{HTML}{8F1D1D}
\definecolor{dsDim}{HTML}{85837B}
\renewcommand{\arraystretch}{1.1}
\begin{tabular}{@{}l l@{\ \ }r@{\hspace{4pt}}r@{\hspace{4pt}}r r@{{\scriptsize\color{dsDim}\ $\pm$\ }}l c l@{}}
\toprule
\multicolumn{5}{l}{} & \multicolumn{3}{c}{Time (ms)\ {\scriptsize\color{dsDim}median of 10 runs}} & \\
\cmidrule(lr){6-8}
\multicolumn{1}{@{}c}{Encoding} & \multicolumn{4}{c}{\makecell{Solved\\{\scriptsize\color{dsDim}\textcolor{dsSat}{\rule{0.8ex}{0.8ex}} SAT\enspace \textcolor{dsUnsat}{\rule{0.8ex}{0.8ex}} UNSAT\enspace \textcolor{dsTimeout}{\rule{0.8ex}{0.8ex}} timeout}}} & \multicolumn{2}{c}{Mean {\scriptsize\color{dsDim}$\pm$ SD}} & \multicolumn{1}{c}{\makecell{Median\\instance}} & \multicolumn{1}{c@{}}{\makecell{Instances solved\\{\scriptsize\color{dsDim}by time, mean of 10 runs}}} \\
\midrule
\textbf{LLM-Synthesized} & \multicolumn{4}{l}{{\scriptsize\color{dsDim}100 instances}} & \multicolumn{2}{c}{} & & \raisebox{-\dimexpr 5.83\normalbaselineskip+2pt\relax}[0pt][0pt]{\includegraphics[height=\dimexpr 6.6\normalbaselineskip+2pt\relax]{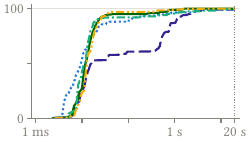}} \\
\addlinespace[2pt]
\textcolor[HTML]{35248F}{\rule[0.5ex]{6pt}{1pt}\hspace{2pt}\rule[0.5ex]{5pt}{1pt}}\ \naive & \textcolor{dsSat}{\rule{0.701cm}{1.3ex}}\hspace{0.6pt}\textcolor{dsUnsat}{\rule{0.495cm}{1.3ex}}\hspace{0.6pt}\textcolor{dsTimeout}{\rule{0.040cm}{1.3ex}} & 58 & 41 & 1 & 578 & {\scriptsize\color{dsDim}2068} & \hphantom{\textbf{384.8}}\llap{15.8} & \\
\textcolor[HTML]{2A78D6}{\rule[0.5ex]{1pt}{1pt}\hspace{1.3pt}\rule[0.5ex]{1pt}{1pt}\hspace{1.3pt}\rule[0.5ex]{1pt}{1pt}\hspace{1.3pt}\rule[0.5ex]{1pt}{1pt}\hspace{1.3pt}\rule[0.5ex]{1pt}{1pt}\hspace{1.3pt}\rule[0.5ex]{1pt}{1pt}\hspace{0.5pt}}\ \epoch & \textcolor{dsSat}{\rule{0.701cm}{1.3ex}}\hspace{0.6pt}\textcolor{dsUnsat}{\rule{0.495cm}{1.3ex}}\hspace{0.6pt}\textcolor{dsTimeout}{\rule{0.040cm}{1.3ex}} & 58 & 41 & 1 & 304 & {\scriptsize\color{dsDim}2047} & \hphantom{\textbf{384.8}}\llap{\underline{\textbf{9.5}}} & \\
\textcolor[HTML]{006400}{\rule[0.5ex]{13pt}{1pt}}\ \hybridymd & \textcolor{dsSat}{\rule{0.725cm}{1.3ex}}\hspace{0.6pt}\textcolor{dsUnsat}{\rule{0.504cm}{1.3ex}} & \underline{\textbf{59}} & \underline{\textbf{41}} & 0 & 50 & {\scriptsize\color{dsDim}191} & \hphantom{\textbf{384.8}}\llap{11.8} & \\
\textcolor[HTML]{1BAF7A}{\rule[0.5ex]{4pt}{1pt}\hspace{1.3pt}\rule[0.5ex]{1pt}{1pt}\hspace{1.3pt}\rule[0.5ex]{4pt}{1pt}\hspace{1.3pt}\rule[0.5ex]{0.1pt}{1pt}}\ \hybridepoch & \textcolor{dsSat}{\rule{0.725cm}{1.3ex}}\hspace{0.6pt}\textcolor{dsUnsat}{\rule{0.504cm}{1.3ex}} & \underline{\textbf{59}} & \underline{\textbf{41}} & 0 & 257 & {\scriptsize\color{dsDim}1236} & \hphantom{\textbf{384.8}}\llap{11.0} & \\
\textcolor[HTML]{F0AC0C}{\rule[0.5ex]{4pt}{1pt}\hspace{1.3pt}\rule[0.5ex]{1pt}{1pt}\hspace{1.3pt}\rule[0.5ex]{1pt}{1pt}\hspace{1.3pt}\rule[0.5ex]{3.1pt}{1pt}}\ \ab & \textcolor{dsSat}{\rule{0.725cm}{1.3ex}}\hspace{0.6pt}\textcolor{dsUnsat}{\rule{0.504cm}{1.3ex}} & \underline{\textbf{59}} & \underline{\textbf{41}} & 0 & \underline{\textbf{40}} & {\scriptsize\color{dsDim}168} & \hphantom{\textbf{384.8}}\llap{12.7} & \\
\midrule[\heavyrulewidth]
\textbf{Grammar-Sampled} & \multicolumn{4}{l}{{\scriptsize\color{dsDim}500 instances}} & \multicolumn{2}{c}{} & & \raisebox{-\dimexpr 5.83\normalbaselineskip+2pt\relax}[0pt][0pt]{\includegraphics[height=\dimexpr 6.6\normalbaselineskip+2pt\relax]{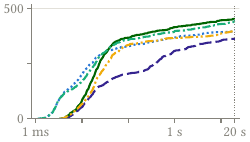}} \\
\addlinespace[2pt]
\textcolor[HTML]{35248F}{\rule[0.5ex]{6pt}{1pt}\hspace{2pt}\rule[0.5ex]{5pt}{1pt}}\ \naive & \textcolor{dsSat}{\rule{0.271cm}{1.3ex}}\hspace{0.6pt}\textcolor{dsUnsat}{\rule{0.606cm}{1.3ex}}\hspace{0.6pt}\textcolor{dsTimeout}{\rule{0.331cm}{1.3ex}} & 112 & 251 & 137 & 6139 & {\scriptsize\color{dsDim}8764} & \hphantom{\textbf{384.8}}\llap{384.8} & \\
\textcolor[HTML]{2A78D6}{\rule[0.5ex]{1pt}{1pt}\hspace{1.3pt}\rule[0.5ex]{1pt}{1pt}\hspace{1.3pt}\rule[0.5ex]{1pt}{1pt}\hspace{1.3pt}\rule[0.5ex]{1pt}{1pt}\hspace{1.3pt}\rule[0.5ex]{1pt}{1pt}\hspace{1.3pt}\rule[0.5ex]{1pt}{1pt}\hspace{0.5pt}}\ \epoch & \textcolor{dsSat}{\rule{0.271cm}{1.3ex}}\hspace{0.6pt}\textcolor{dsUnsat}{\rule{0.691cm}{1.3ex}}\hspace{0.6pt}\textcolor{dsTimeout}{\rule{0.246cm}{1.3ex}} & 112 & 286 & 102 & 4432 & {\scriptsize\color{dsDim}8091} & \hphantom{\textbf{384.8}}\llap{\underline{\textbf{15.7}}} & \\
\textcolor[HTML]{006400}{\rule[0.5ex]{13pt}{1pt}}\ \hybridymd & \textcolor{dsSat}{\rule{0.331cm}{1.3ex}}\hspace{0.6pt}\textcolor{dsUnsat}{\rule{0.763cm}{1.3ex}}\hspace{0.6pt}\textcolor{dsTimeout}{\rule{0.114cm}{1.3ex}} & \underline{\textbf{137}} & \underline{\textbf{316}} & 47 & \underline{\textbf{2400}} & {\scriptsize\color{dsDim}6057} & \hphantom{\textbf{384.8}}\llap{23.8} & \\
\textcolor[HTML]{1BAF7A}{\rule[0.5ex]{4pt}{1pt}\hspace{1.3pt}\rule[0.5ex]{1pt}{1pt}\hspace{1.3pt}\rule[0.5ex]{4pt}{1pt}\hspace{1.3pt}\rule[0.5ex]{0.1pt}{1pt}}\ \hybridepoch & \textcolor{dsSat}{\rule{0.309cm}{1.3ex}}\hspace{0.6pt}\textcolor{dsUnsat}{\rule{0.761cm}{1.3ex}}\hspace{0.6pt}\textcolor{dsTimeout}{\rule{0.138cm}{1.3ex}} & 128 & 315 & 57 & 3027 & {\scriptsize\color{dsDim}6709} & \hphantom{\textbf{384.8}}\llap{19.4} & \\
\textcolor[HTML]{F0AC0C}{\rule[0.5ex]{4pt}{1pt}\hspace{1.3pt}\rule[0.5ex]{1pt}{1pt}\hspace{1.3pt}\rule[0.5ex]{1pt}{1pt}\hspace{1.3pt}\rule[0.5ex]{3.1pt}{1pt}}\ \ab & \textcolor{dsSat}{\rule{0.263cm}{1.3ex}}\hspace{0.6pt}\textcolor{dsUnsat}{\rule{0.701cm}{1.3ex}}\hspace{0.6pt}\textcolor{dsTimeout}{\rule{0.244cm}{1.3ex}} & 109 & 290 & 101 & 4663 & {\scriptsize\color{dsDim}8135} & \hphantom{\textbf{384.8}}\llap{36.1} & \\
\midrule[\heavyrulewidth]
\textbf{Legally Grounded} & \multicolumn{4}{l}{{\scriptsize\color{dsDim}200 instances}} & \multicolumn{2}{c}{} & & \raisebox{-\dimexpr 5.83\normalbaselineskip+2pt\relax}[0pt][0pt]{\includegraphics[height=\dimexpr 6.6\normalbaselineskip+2pt\relax]{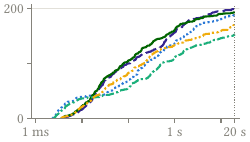}} \\
\addlinespace[2pt]
\textcolor[HTML]{35248F}{\rule[0.5ex]{6pt}{1pt}\hspace{2pt}\rule[0.5ex]{5pt}{1pt}}\ \naive & \textcolor{dsSat}{\rule{1.223cm}{1.3ex}}\hspace{0.6pt}\textcolor{dsTimeout}{\rule{0.040cm}{1.3ex}} & \underline{\textbf{199}} & \underline{\textbf{0}} & 1 & \underline{\textbf{1074}} & {\scriptsize\color{dsDim}2754} & \hphantom{\textbf{384.8}}\llap{102.2} & \\
\textcolor[HTML]{2A78D6}{\rule[0.5ex]{1pt}{1pt}\hspace{1.3pt}\rule[0.5ex]{1pt}{1pt}\hspace{1.3pt}\rule[0.5ex]{1pt}{1pt}\hspace{1.3pt}\rule[0.5ex]{1pt}{1pt}\hspace{1.3pt}\rule[0.5ex]{1pt}{1pt}\hspace{1.3pt}\rule[0.5ex]{1pt}{1pt}\hspace{0.5pt}}\ \epoch & \textcolor{dsSat}{\rule{1.155cm}{1.3ex}}\hspace{0.6pt}\textcolor{dsTimeout}{\rule{0.074cm}{1.3ex}} & 188 & 0 & 12 & 2473 & {\scriptsize\color{dsDim}5315} & \hphantom{\textbf{384.8}}\llap{242.5} & \\
\textcolor[HTML]{006400}{\rule[0.5ex]{13pt}{1pt}}\ \hybridymd & \textcolor{dsSat}{\rule{1.186cm}{1.3ex}}\hspace{0.6pt}\textcolor{dsTimeout}{\rule{0.043cm}{1.3ex}} & 193 & 0 & 7 & 1477 & {\scriptsize\color{dsDim}4200} & \hphantom{\textbf{384.8}}\llap{\underline{\textbf{78.2}}} & \\
\textcolor[HTML]{1BAF7A}{\rule[0.5ex]{4pt}{1pt}\hspace{1.3pt}\rule[0.5ex]{1pt}{1pt}\hspace{1.3pt}\rule[0.5ex]{4pt}{1pt}\hspace{1.3pt}\rule[0.5ex]{0.1pt}{1pt}}\ \hybridepoch & \textcolor{dsSat}{\rule{0.940cm}{1.3ex}}\hspace{0.6pt}\textcolor{dsTimeout}{\rule{0.289cm}{1.3ex}} & 153 & 0 & 47 & 5935 & {\scriptsize\color{dsDim}8413} & \hphantom{\textbf{384.8}}\llap{525.9} & \\
\textcolor[HTML]{F0AC0C}{\rule[0.5ex]{4pt}{1pt}\hspace{1.3pt}\rule[0.5ex]{1pt}{1pt}\hspace{1.3pt}\rule[0.5ex]{1pt}{1pt}\hspace{1.3pt}\rule[0.5ex]{3.1pt}{1pt}}\ \ab & \textcolor{dsSat}{\rule{1.045cm}{1.3ex}}\hspace{0.6pt}\textcolor{dsTimeout}{\rule{0.184cm}{1.3ex}} & 170 & 0 & 30 & 4383 & {\scriptsize\color{dsDim}7517} & \hphantom{\textbf{384.8}}\llap{212.0} & \\
\bottomrule
\end{tabular}

\end{table}

Table~\ref{tab:solver-outcomes} shows the solvers' evaluation result. \emph{Solved} gives how many instances the encoding solves within the time limit, as SAT and as UNSAT, and how many it times out on, each averaged over the 10 runs. The two time columns give the encoding's \emph{Mean} time per instance, with a timed-out instance counting as 20\,s, and its time on the \emph{Median instance}. We compute both on each run separately and report their median over the 10 runs. Beside the mean, after $\pm$, is the standard deviation of the time per instance in the runs that give the median mean. The last column plots, for each encoding, the number of instances it solves within a given time, averaged over the 10 runs, so each curve ends at the sum of the encoding's SAT and UNSAT counts. Its time axis is logarithmic, so the horizontal gap between two curves shows how many times slower one encoding is than another at the same number of solved instances.

On the LLM-synthesized benchmark, every encoding solves at least 99 of the 100 instances: the encodings differ mainly in speed. \naive{} is the slowest, with the highest mean and median instance times: its curve stays within about a factor of two of the leading curve for the first half of the instances, but then falls up to 61 times behind. \ab{} has the lowest mean times, and its and \hybridymd{}'s small standard deviations show that none of their instances takes long. The grammar-sampled benchmark is the hardest: every encoding times out on 47 to 137 of its 500 instances. Since each timeout counts as 20\,s, the timeouts dominate the mean times, which rank the encodings almost exactly as their timeouts do. The curves show that \epoch{} and \hybridepoch{} solve the most instances at first, which gives them the low median instance times. After a few tens of milliseconds, however, \hybridymd{} overtakes them and keeps the lead until the timeout: it times out the least, solves the most instances, and has the lowest mean time. On the legally grounded benchmark, \epoch{} and \hybridepoch{} again lead at first, but time out on 12 and 47 instances. \naive{} times out the least (1), and \hybridymd{} on 7; their curves stay close, and \naive{} has the lowest mean time and \hybridymd{} the lowest median instance time.

No encoding is therefore best overall. On the grammar-sampled and legally grounded benchmarks, the encodings that lead early are not the ones that solve the most instances, and on the LLM-synthesized benchmark, \hybridepoch{} solves every instance yet has five times the mean time of \hybridymd{}. Still, every encoding is the fastest on some constraints: of the 768 constraints that at least one encoding solves, \epoch{} is the fastest on 350, \hybridymd{} on 154, \hybridepoch{} on 107, \ab{} on 85, and even \naive{} on 72. \textbf{Each encoding thus has its own strengths.}

Looking deeper into the results with the features of Sec.~\ref{sec:dataset}, we find that each encoding's performance is driven mainly by the operation it encodes most expensively (Sec.~\ref{sec:solver}). The features count how often a constraint uses that operation, which largely decides how well the encoding does.

\textbf{Day additions are what the advanced encodings speed up.} \naive{} encodes adding $n$ days as $n$ nested ITEs (Sec.~\ref{sec:naive}): it never times out on the 392 constraints that add no days, but times out on 130 of the 190 that add over 365 days in total, 
where it is on average $47\times$ slower than \epoch{} with its geometric mean over the constraints.
Legal constraints mostly shift dates by whole years (65\% of their date--period operations), and 75\% of them add no days, so \naive{} stays competitive there: it times out the least, on one constraint per run, which adds 133 or 1,080 days, and has the lowest mean time, although \hybridymd{} is still faster on the median instance.

\textbf{\epoch{} and \ab{} each handle one kind of arithmetic well.} \epoch{} adds days directly to the offset but must convert it by integer division to read a date as $\Date{y,m,d}$, e.g., to add months (Sec.~\ref{sec:epoch}): on the 169 constraints that need no such date, it never times out, and it or \hybridepoch{} is the fastest on 160. \ab{} adds months directly to its month count but splits on every day addition (Sec.~\ref{sec:alpha-beta}): on the 216 constraints that only add years or months, it never times out, while \epoch{} and \hybridepoch{} time out on 36 and 41, but with day additions it is on average $3.6\times$ slower than \hybridymd{}.

\textbf{\hybrid pays off when both representations needed.} On the 482 constraints that need only one representation, a \hybrid variant that starts in it needs no conversion and has nothing to gain: \hybridepoch{} is on average $1.2\times$ slower than \epoch{}. On the 235 that need at least two conversions under either variant, both \hybrid variants are on average faster than every single-representation encoding, by at least $1.27\times$ over \epoch{}, and time out the least, on 47 and 45 constraints against 68 to 102. The better of the two is the one initialized in the representation its variables need (Sec.~\ref{sec:hybrid}): \hybridymd{}, initialized as the triple $(a_y, a_m, a_d)$, is faster on 287 of the 396 constraints where more variables need only the triple, and \hybridepoch{}, initialized as the offset $a_\Delta$, on 104 of the 127 where more need only the offset. Together, these explain why \hybridymd{} times out the least overall, on 54 constraints against 104 to 139 for the other encodings.

To leverage these strengths, we build a \emph{router} that picks an encoding for each constraint at solve time, inspired by per-instance algorithm selection in SATzilla~\cite{satzilla} for SAT and MachSMT~\cite{machsmt} for SMT. For every pair of encodings, a random forest predicts which of the two would solve faster, using its 61 static features (Sec.~\ref{sec:dataset}); each training example is weighted by the time difference between the two encodings, so that the forests focus on the choices that matter most, such as avoiding timeout. The router picks the encoding that wins the most pairs, and falls back to \hybridymd{} when the vote is tied or its winner is not predicted to beat \hybridymd{}. We evaluate it by 5-fold cross-validation stratified by benchmark, with a new random split in each of the 10 runs: \datesat{} solves every constraint with the router of its fold, which was not trained on it, and the router's time includes extracting the features and querying the forests. We compare the router with always using one encoding and with an ideal \emph{oracle}, which picks the fastest encoding for every constraint and so provides an upper bound for the router.
\begin{figure}[t]
\centering
\includegraphics[width=\linewidth]{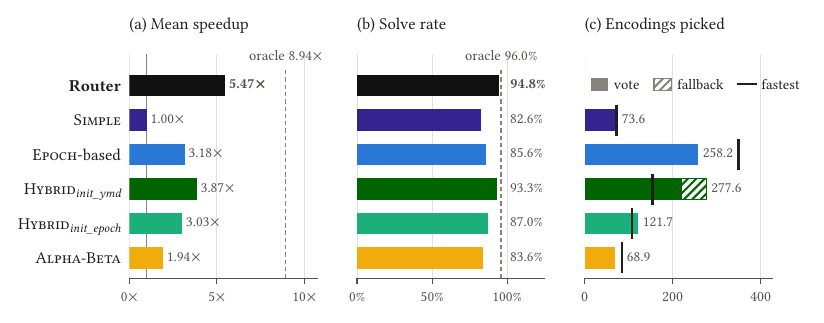}
\caption{The router against always using one encoding across all of \dataset{}: (a) geometric-mean per-constraint speedup over \naive{}, (b) share of constraints solved within the timeout, and (c) how often the router picks each encoding, by vote or by falling back to \hybridymd{}. In (c), the marker labelled \emph{fastest} shows how many constraints each encoding is the fastest on, which is how often the oracle picks it, so the gap between a bar and its marker shows how far the router's mix of encodings is from the oracle's.}
\label{fig:router}
\end{figure}
As Fig.~\ref{fig:router} shows, the router outperforms every single encoding on both measures: its mean speedup over \naive{}, the geometric mean over the constraints, is $5.47\times$, against $3.87\times$ for \hybridymd{}, the best single encoding, and it solves 94.8\% of the constraints, against 93.3\%. Only in median speedup, which reflects the typical constraint, is it third, at $2.32\times$ against $2.54\times$ for \hybridepoch{} and $2.53\times$ for \epoch{}: routing at solve time adds about 2\,ms to every constraint, which counts for little in total but weighs on the half of the constraints that are solved in under 20\,ms.
The router uses all encodings (Fig.~\ref{fig:router}c) but favors \hybridymd{}, the encoding that times out least, picking it on 278 constraints, 57 of them by falling back when its vote is uncertain; \epoch{}, the fastest on 350, was picked by the router on 258 instances. Its pick is the fastest encoding on 358 of the 768 constraints that some encoding solves, and when it is not, it is typically only $1.38\times$ slower than the fastest. 
Overall, choosing an encoding dynamically for each constraint turns the encodings' different strengths into a solver that is faster on average and solves more constraints than any single encoding.

\section{Threats to Validity and Mitigations Thereof}
\label{sec:discussion}

Construct validity depends on our choice of semantics, encodings, or their implementation.
To check that the semantics of Sec.~\ref{sec:semantics} follows the calendar convention it claims, we test its Verus specification, through an executable implementation verified to agree with it, against Python's \code{datetime} and \code{dateutil} and Java's \code{java.time}, on date validity, date comparison, and date-period arithmetic.\footnote{Python's \code{datetime} only represents years 1--9999; within that range we check date validity exhaustively, and date comparison and date-period arithmetic on 1.5 million sampled cases. We also compare with \code{java.time} on 3.5 million sampled cases whose dates are drawn from three ranges: years 1--9999, years $-50{,}000$ to $0$, and years $10{,}000$ to $50{,}000$.}
Sec.~\ref{sec:correctness} proves every encoding equisatisfiable with this semantics. The implementation is tested by a suite of about 3,500 unit tests covering each encoding, along with property-based tests of the core date and period operations.
On \dataset{}, models are checked by evaluating each constraint on the returned assignment. Date validity, ordering and date-period addition are computed by an executable implementation that is verified in Verus to agree with the reference semantics of Sec.~\ref{sec:semantics}; Boolean and integer subterms are evaluated in Python. We also perform differential testing that all encodings that solve an instance agree on whether it is SAT or UNSAT.
Our results reflect one implementation of the encodings and one backend solver, Z3 4.16.0.

To mitigate threats to internal validity from timing variability and biased comparisons, we evaluate all encodings on the same constraints and isolated server, with the same Python and Z3 versions and a common 20\,s timeout, repeating each experiment 10 times (Sec.~\ref{sec:evaluation}). For the router, 5-fold cross-validation keeps each evaluated constraint out of its model's training set, and reported times include feature extraction and model queries. Results remain sensitive to the timeout: instances that exceed it contribute 20\,s to the reported mean, which therefore does not capture their full solving cost. To mitigate threats to external validity from benchmark construction choices, we combine three complementary sources targeting calendar edge cases, varied constraint structures, and legally motivated reasoning tasks (Sec.~\ref{sec:dataset}). This diversity broadens the operations and structures exercised, but we cannot guarantee that the results will generalize to other benchmarks.

\section{Related Work}
\label{sec:related}

Our work is influenced by past work in developing specialized support for strings in SMT solvers~\cite{Kiezun09, Saxena10, Zheng13, Liang14, Trinh14, Berzish17, Zheng17, Mora21} based on domain-specific insights as well as applications to symbolic execution of programs that manipulate strings. 

To the best of our knowledge, no explicit support exists for date and calendar logic in SMT solvers today. The main applications of constraint solving to temporal values in the literature are in modeling real-time systems~\cite{Dutertre04} or optimizing job scheduling across machines~\cite{Baptiste01-ConstraintScheduling, Xiong22-JSSP}; these techniques do not deal with absolute date representations.
While efficient algorithms exist for performing date and period arithmetic on concrete values~\cite{Fliegel68, Reingold18, Hinnant13}, prior work has not addressed solving constraints involving symbolic dates and periods.

We build upon prior work on formalizing the semantics of date and period arithmetic~\cite{Monat24}. CUTECat~\cite{Goutagny25} performs symbolic execution of programs written in a domain-specific language for legal analysis. This tool needs to reason about dates (similar to our example in Sec.~\ref{sec:motivation}, Fig.~\ref{fig:motivating-example-legal}); however, the paper indicates that due to the complexity of date semantics, they concretize date values at the cost of completeness. Our work directly addresses this gap.

Our router, which picks an encoding for each constraint as it is solved, is inspired by the common practice of per-instance algorithm selection~\cite{Rice76, Kerschke19}, in which a portfolio of specialized solvers is combined with a learned selector that picks one from static features of the instance, as in SATzilla~\cite{satzilla}, MachSMT~\cite{machsmt}, and MedleySolver~\cite{medleysolver}. The same idea underlies recent work that uses coding agents to synthesize the specialists themselves, from per-family SAT solvers~\cite{Green26-Hyperspecialization} to algorithms for systems problems with an automated verifier~\cite{Cheng25-Barbarians, Novikov25-AlphaEvolve}.

More recently, the research community has begun to curate datasets of date- and time-related bugs in software~\cite{Tiwari25}, as well as benchmarks aimed at evaluating the temporal reasoning capabilities of LLMs~\cite{Fatemi24, Gaere25}. Although these efforts do not directly target constraint solvability or program verification, they underscore the growing prevalence and practical importance of date/time logic in modern software systems. We hope that \dataset{} can serve as a foundation for future work on principled reasoning, analysis, and verification of programs involving temporal computations.
\section{Conclusion}
\label{sec:conclusion}
Reasoning symbolically about dates and calendar periods is tricky, and until now no SMT solver or verification tool has supported it natively. We introduced \datesat{}, the first framework for expressing and solving constraints involving date and period arithmetic. We formalized its semantics, presented four encoding strategies that reduce \datesat{} to linear integer arithmetic with mechanized proofs of their equisatisfiability, and evaluated them on \dataset{}, a benchmark of \totalConstraints{} constraints, finding that the encodings have complementary strengths that a lightweight router can take advantage of by picking an encoding dynamically for each constraint. The ideas behind \datesat{} are being incorporated into Python's Hypothesis and CrossHair tools, bringing symbolic reasoning about dates to property-based testing and symbolic execution respectively. We expect similar benefits for other applications, from program verification to agentic AI systems that must reason accurately about dates.

\begin{acks}
This work was supported in large part by \grantsponsor{nsf}{National Science Foundation}{https://www.nsf.gov} Grant \href{https://www.nsf.gov/awardsearch/showAward?AWD_ID=2429384}{\textcolor{black}{CCF-2429384}}.
\end{acks}

\bibliographystyle{ACM-Reference-Format}
\bibliography{bib}


\begin{thebibliography}{49}


\ifx \showCODEN    \undefined \def \showCODEN     #1{\unskip}     \fi
\ifx \showISBNx    \undefined \def \showISBNx     #1{\unskip}     \fi
\ifx \showISBNxiii \undefined \def \showISBNxiii  #1{\unskip}     \fi
\ifx \showISSN     \undefined \def \showISSN      #1{\unskip}     \fi
\ifx \showLCCN     \undefined \def \showLCCN      #1{\unskip}     \fi
\ifx \shownote     \undefined \def \shownote      #1{#1}          \fi
\ifx \showarticletitle \undefined \def \showarticletitle #1{#1}   \fi
\ifx \showURL      \undefined \def \showURL       {\relax}        \fi
\providecommand\bibfield[2]{#2}
\providecommand\bibinfo[2]{#2}
\providecommand\natexlab[1]{#1}
\providecommand\showeprint[2][]{arXiv:#2}

\bibitem[Azure(2012)]%
        {Laing12}
\bibfield{author}{\bibinfo{person}{Microsoft Azure}.} \bibinfo{year}{2012}\natexlab{}.
\newblock \bibinfo{title}{Summary of {Windows Azure} Service Disruption on {Feb 29th, 2012}}.
\newblock \bibinfo{howpublished}{\url{https://azure.microsoft.com/en-us/blog/summary-of-windows-azure-service-disruption-on-feb-29th-2012/}}.
\newblock
\shownote{Retrieved Apr 26, 2026}.
\newblock


\bibitem[Baptiste et~al\mbox{.}(2001)]%
        {Baptiste01-ConstraintScheduling}
\bibfield{author}{\bibinfo{person}{Philippe Baptiste}, \bibinfo{person}{Claude Le~Pape}, {and} \bibinfo{person}{Wim Nuijten}.} \bibinfo{year}{2001}\natexlab{}.
\newblock \bibinfo{booktitle}{\emph{Constraint-based scheduling: applying constraint programming to scheduling problems}}. Vol.~\bibinfo{volume}{39}.
\newblock \bibinfo{publisher}{Springer Science \& Business Media}.
\newblock


\bibitem[Barrett and Tinelli(2018)]%
        {Barrett18}
\bibfield{author}{\bibinfo{person}{Clark Barrett} {and} \bibinfo{person}{Cesare Tinelli}.} \bibinfo{year}{2018}\natexlab{}.
\newblock \showarticletitle{Satisfiability modulo theories}.
\newblock In \bibinfo{booktitle}{\emph{Handbook of model checking}}. \bibinfo{publisher}{Springer}, \bibinfo{pages}{305--343}.
\newblock


\bibitem[Berzish et~al\mbox{.}(2017)]%
        {Berzish17}
\bibfield{author}{\bibinfo{person}{Murphy Berzish}, \bibinfo{person}{Vijay Ganesh}, {and} \bibinfo{person}{Yunhui Zheng}.} \bibinfo{year}{2017}\natexlab{}.
\newblock \showarticletitle{Z3str3: A string solver with theory-aware heuristics}. In \bibinfo{booktitle}{\emph{2017 Formal Methods in Computer Aided Design (FMCAD)}}. IEEE, \bibinfo{pages}{55--59}.
\newblock


\bibitem[Blair-Stanek et~al\mbox{.}(2023)]%
        {Blair-Stanek23-GPT3Statutory}
\bibfield{author}{\bibinfo{person}{Andrew Blair-Stanek}, \bibinfo{person}{Nils Holzenberger}, {and} \bibinfo{person}{Benjamin Van~Durme}.} \bibinfo{year}{2023}\natexlab{}.
\newblock \showarticletitle{Can {GPT-3} perform statutory reasoning?}
\newblock \bibinfo{journal}{\emph{arXiv preprint arXiv:2302.06100}} (\bibinfo{year}{2023}).
\newblock


\bibitem[Cheian(2022)]%
        {Cheian22}
\bibfield{author}{\bibinfo{person}{Dinis Cheian}.} \bibinfo{year}{2022}\natexlab{}.
\newblock \showarticletitle{I see dead patents: How bugs in the patent system keep expired patents alive}.
\newblock \bibinfo{journal}{\emph{Fordham Intell. Prop. Media \& Ent. LJ}}  \bibinfo{volume}{33} (\bibinfo{year}{2022}), \bibinfo{pages}{1}.
\newblock


\bibitem[Cheng et~al\mbox{.}(2025)]%
        {Cheng25-Barbarians}
\bibfield{author}{\bibinfo{person}{Audrey Cheng}, \bibinfo{person}{Shu Liu}, \bibinfo{person}{Melissa Pan}, \bibinfo{person}{Zhifei Li}, \bibinfo{person}{Bowen Wang}, \bibinfo{person}{Alex Krentsel}, \bibinfo{person}{Tian Xia}, \bibinfo{person}{Mert Cemri}, \bibinfo{person}{Jongseok Park}, \bibinfo{person}{Shuo Yang}, \bibinfo{person}{Jeff Chen}, \bibinfo{person}{Lakshya Agrawal}, \bibinfo{person}{Aditya Desai}, \bibinfo{person}{Jiarong Xing}, \bibinfo{person}{Koushik Sen}, \bibinfo{person}{Matei Zaharia}, {and} \bibinfo{person}{Ion Stoica}.} \bibinfo{year}{2025}\natexlab{}.
\newblock \bibinfo{title}{Barbarians at the Gate: How {AI} is Upending Systems Research}.
\newblock
\showeprint[arxiv]{2510.06189}~[cs.AI]
\urldef\tempurl%
\url{https://arxiv.org/abs/2510.06189}
\showURL{%
\tempurl}


\bibitem[Chistikov(2024)]%
        {Chistikov24}
\bibfield{author}{\bibinfo{person}{Dmitry Chistikov}.} \bibinfo{year}{2024}\natexlab{}.
\newblock \showarticletitle{An introduction to the theory of linear integer arithmetic}. In \bibinfo{booktitle}{\emph{44th IARCS Annual Conference on Foundations of Software Technology and Theoretical Computer Science (FSTTCS 2024)}}, Vol.~\bibinfo{volume}{323}. Schloss Dagstuhl—Leibniz-Zentrum f{\"u}r Informatik, \bibinfo{pages}{1}.
\newblock


\bibitem[De~Moura and Bj\o{}rner(2008)]%
        {Z3}
\bibfield{author}{\bibinfo{person}{Leonardo De~Moura} {and} \bibinfo{person}{Nikolaj Bj\o{}rner}.} \bibinfo{year}{2008}\natexlab{}.
\newblock \showarticletitle{Z3: an efficient {SMT} solver}. In \bibinfo{booktitle}{\emph{Proceedings of the Theory and Practice of Software, 14th International Conference on Tools and Algorithms for the Construction and Analysis of Systems}} (Budapest, Hungary) \emph{(\bibinfo{series}{TACAS'08/ETAPS'08})}. \bibinfo{publisher}{Springer-Verlag}, \bibinfo{address}{Berlin, Heidelberg}, \bibinfo{pages}{337–340}.
\newblock
\showISBNx{3540787992}


\bibitem[Dutertre and Sorea(2004)]%
        {Dutertre04}
\bibfield{author}{\bibinfo{person}{Bruno Dutertre} {and} \bibinfo{person}{Maria Sorea}.} \bibinfo{year}{2004}\natexlab{}.
\newblock \showarticletitle{Modeling and verification of a fault-tolerant real-time startup protocol using calendar automata}. In \bibinfo{booktitle}{\emph{International Symposium on Formal Techniques in Real-Time and Fault-Tolerant Systems}}. Springer, \bibinfo{pages}{199--214}.
\newblock


\bibitem[Fatemi et~al\mbox{.}(2024)]%
        {Fatemi24}
\bibfield{author}{\bibinfo{person}{Bahare Fatemi}, \bibinfo{person}{Mehran Kazemi}, \bibinfo{person}{Anton Tsitsulin}, \bibinfo{person}{Karishma Malkan}, \bibinfo{person}{Jinyeong Yim}, \bibinfo{person}{John Palowitch}, \bibinfo{person}{Sungyong Seo}, \bibinfo{person}{Jonathan Halcrow}, {and} \bibinfo{person}{Bryan Perozzi}.} \bibinfo{year}{2024}\natexlab{}.
\newblock \showarticletitle{Test of time: A benchmark for evaluating {LLM}s on temporal reasoning}.
\newblock \bibinfo{journal}{\emph{arXiv preprint arXiv:2406.09170}} (\bibinfo{year}{2024}).
\newblock


\bibitem[Fliegel and Van~Flandern(1968)]%
        {Fliegel68}
\bibfield{author}{\bibinfo{person}{Henry~F. Fliegel} {and} \bibinfo{person}{Thomas~C. Van~Flandern}.} \bibinfo{year}{1968}\natexlab{}.
\newblock \showarticletitle{A machine algorithm for processing calendar dates}.
\newblock \bibinfo{journal}{\emph{Commun. ACM}} \bibinfo{volume}{11}, \bibinfo{number}{10} (\bibinfo{year}{1968}), \bibinfo{pages}{657}.
\newblock
\href{https://doi.org/10.1145/364096.364097}{doi:\nolinkurl{10.1145/364096.364097}}


\bibitem[Gaere and Wangenheim(2025)]%
        {Gaere25}
\bibfield{author}{\bibinfo{person}{Edward Gaere} {and} \bibinfo{person}{Florian Wangenheim}.} \bibinfo{year}{2025}\natexlab{}.
\newblock \showarticletitle{{DATETIME}: A new benchmark to measure {LLM} translation and reasoning capabilities}.
\newblock \bibinfo{journal}{\emph{arXiv preprint arXiv:2504.16155}} (\bibinfo{year}{2025}).
\newblock


\bibitem[Gelbmann(2026)]%
        {Gelbmann26}
\bibfield{author}{\bibinfo{person}{Lukas Gelbmann}.} \bibinfo{year}{2026}\natexlab{}.
\newblock \bibinfo{title}{A glitch in {February} of the year 0}.
\newblock \bibinfo{howpublished}{\url{https://28times.com/blog/2026-06-26-february-of-the-year-0}}.
\newblock
\shownote{Retrieved Oct 4, 2026}.
\newblock


\bibitem[Goebel et~al\mbox{.}(2023)]%
        {Goebel23-COLIEE23}
\bibfield{author}{\bibinfo{person}{Randy Goebel}, \bibinfo{person}{Yoshinobu Kano}, \bibinfo{person}{Mi-Young Kim}, \bibinfo{person}{Juliano Rabelo}, \bibinfo{person}{Ken Satoh}, {and} \bibinfo{person}{Masaharu Yoshioka}.} \bibinfo{year}{2023}\natexlab{}.
\newblock \showarticletitle{Summary of the competition on legal information, extraction/entailment ({COLIEE}) 2023}. In \bibinfo{booktitle}{\emph{Proceedings of the Nineteenth International Conference on Artificial Intelligence and Law}}. \bibinfo{pages}{472--480}.
\newblock


\bibitem[Goutagny et~al\mbox{.}(2025)]%
        {Goutagny25}
\bibfield{author}{\bibinfo{person}{Pierre Goutagny}, \bibinfo{person}{Aymeric Fromherz}, {and} \bibinfo{person}{Rapha{\"e}l Monat}.} \bibinfo{year}{2025}\natexlab{}.
\newblock \showarticletitle{{CUTECat}: Concolic Execution for Computational Law}. In \bibinfo{booktitle}{\emph{European Symposium on Programming}}. Springer, \bibinfo{pages}{31--61}.
\newblock


\bibitem[{Government Publishing Office}(2018)]%
        {irc-section121}
\bibfield{author}{\bibinfo{person}{{Government Publishing Office}}.} \bibinfo{year}{2018}\natexlab{}.
\newblock \showarticletitle{{Internal Revenue Code}}.
\newblock \bibinfo{journal}{\emph{Code of Laws of the United States of America}} \bibinfo{volume}{26}, \bibinfo{number}{121} (\bibinfo{year}{2018}).
\newblock


\bibitem[Green et~al\mbox{.}(2026)]%
        {Green26-Hyperspecialization}
\bibfield{author}{\bibinfo{person}{Harrison Green}, \bibinfo{person}{Claire {Le Goues}}, {and} \bibinfo{person}{Fraser Brown}.} \bibinfo{year}{2026}\natexlab{}.
\newblock \bibinfo{title}{The Case for Automated Hyperspecialization: Evidence from {SAT}}.
\newblock
\showeprint[arxiv]{2609.14836}~[cs.SE]
\urldef\tempurl%
\url{https://arxiv.org/abs/2609.14836}
\showURL{%
\tempurl}


\bibitem[Groce et~al\mbox{.}(2012)]%
        {SwarmTesting}
\bibfield{author}{\bibinfo{person}{Alex Groce}, \bibinfo{person}{Chaoqiang Zhang}, \bibinfo{person}{Eric Eide}, \bibinfo{person}{Yang Chen}, {and} \bibinfo{person}{John Regehr}.} \bibinfo{year}{2012}\natexlab{}.
\newblock \showarticletitle{Swarm testing}. In \bibinfo{booktitle}{\emph{Proceedings of the 2012 International Symposium on Software Testing and Analysis}} (Minneapolis, MN, USA) \emph{(\bibinfo{series}{ISSTA 2012})}. \bibinfo{publisher}{Association for Computing Machinery}, \bibinfo{address}{New York, NY, USA}, \bibinfo{pages}{78–88}.
\newblock
\showISBNx{9781450314541}
\href{https://doi.org/10.1145/2338965.2336763}{doi:\nolinkurl{10.1145/2338965.2336763}}


\bibitem[Guha et~al\mbox{.}(2023)]%
        {Guha24-LegalBench}
\bibfield{author}{\bibinfo{person}{Neel Guha}, \bibinfo{person}{Julian Nyarko}, \bibinfo{person}{Daniel Ho}, \bibinfo{person}{Christopher R{\'e}}, \bibinfo{person}{Adam Chilton}, \bibinfo{person}{Alex Chohlas-Wood}, \bibinfo{person}{Austin Peters}, \bibinfo{person}{Brandon Waldon}, \bibinfo{person}{Daniel Rockmore}, \bibinfo{person}{Diego Zambrano}, {et~al\mbox{.}}} \bibinfo{year}{2023}\natexlab{}.
\newblock \showarticletitle{Legalbench: A collaboratively built benchmark for measuring legal reasoning in large language models}.
\newblock \bibinfo{journal}{\emph{Advances in neural information processing systems}}  \bibinfo{volume}{36} (\bibinfo{year}{2023}), \bibinfo{pages}{44123--44279}.
\newblock


\bibitem[Hinnant(2013)]%
        {Hinnant13}
\bibfield{author}{\bibinfo{person}{Howard Hinnant}.} \bibinfo{year}{2013}\natexlab{}.
\newblock \bibinfo{title}{\texttt{chrono}-Compatible Low-Level Date Algorithms}.
\newblock \bibinfo{howpublished}{\url{https://howardhinnant.github.io/date_algorithms.html}}.
\newblock


\bibitem[Kerschke et~al\mbox{.}(2019)]%
        {Kerschke19}
\bibfield{author}{\bibinfo{person}{Pascal Kerschke}, \bibinfo{person}{Holger~H. Hoos}, \bibinfo{person}{Frank Neumann}, {and} \bibinfo{person}{Heike Trautmann}.} \bibinfo{year}{2019}\natexlab{}.
\newblock \showarticletitle{Automated Algorithm Selection: Survey and Perspectives}.
\newblock \bibinfo{journal}{\emph{Evolutionary Computation}} \bibinfo{volume}{27}, \bibinfo{number}{1} (\bibinfo{year}{2019}), \bibinfo{pages}{3--45}.
\newblock
\href{https://doi.org/10.1162/evco_a_00242}{doi:\nolinkurl{10.1162/evco_a_00242}}


\bibitem[Kiezun et~al\mbox{.}(2009)]%
        {Kiezun09}
\bibfield{author}{\bibinfo{person}{Adam Kiezun}, \bibinfo{person}{Vijay Ganesh}, \bibinfo{person}{Philip~J Guo}, \bibinfo{person}{Pieter Hooimeijer}, {and} \bibinfo{person}{Michael~D Ernst}.} \bibinfo{year}{2009}\natexlab{}.
\newblock \showarticletitle{HAMPI: a solver for string constraints}. In \bibinfo{booktitle}{\emph{Proceedings of the eighteenth international symposium on Software testing and analysis}}. \bibinfo{pages}{105--116}.
\newblock


\bibitem[Lattuada et~al\mbox{.}(2023)]%
        {Lattuada23-Verus}
\bibfield{author}{\bibinfo{person}{Andrea Lattuada}, \bibinfo{person}{Travis Hance}, \bibinfo{person}{Chanhee Cho}, \bibinfo{person}{Matthias Brun}, \bibinfo{person}{Isitha Subasinghe}, \bibinfo{person}{Yi Zhou}, \bibinfo{person}{Jon Howell}, \bibinfo{person}{Bryan Parno}, {and} \bibinfo{person}{Chris Hawblitzel}.} \bibinfo{year}{2023}\natexlab{}.
\newblock \showarticletitle{Verus: Verifying {Rust} programs using linear ghost types}.
\newblock \bibinfo{journal}{\emph{Proceedings of the ACM on Programming Languages}} \bibinfo{volume}{7}, \bibinfo{number}{OOPSLA1} (\bibinfo{year}{2023}), \bibinfo{pages}{286--315}.
\newblock


\bibitem[Liang et~al\mbox{.}(2014)]%
        {Liang14}
\bibfield{author}{\bibinfo{person}{Tianyi Liang}, \bibinfo{person}{Andrew Reynolds}, \bibinfo{person}{Cesare Tinelli}, \bibinfo{person}{Clark Barrett}, {and} \bibinfo{person}{Morgan Deters}.} \bibinfo{year}{2014}\natexlab{}.
\newblock \showarticletitle{A DPLL (T) theory solver for a theory of strings and regular expressions}. In \bibinfo{booktitle}{\emph{International Conference on Computer Aided Verification}}. Springer, \bibinfo{pages}{646--662}.
\newblock


\bibitem[MacIver et~al\mbox{.}(2019)]%
        {MacIver2019Hypothesis}
\bibfield{author}{\bibinfo{person}{David~R MacIver}, \bibinfo{person}{Zac Hatfield-Dodds}, {et~al\mbox{.}}} \bibinfo{year}{2019}\natexlab{}.
\newblock \showarticletitle{Hypothesis: A new approach to property-based testing}.
\newblock \bibinfo{journal}{\emph{Journal of Open Source Software}} \bibinfo{volume}{4}, \bibinfo{number}{43} (\bibinfo{year}{2019}), \bibinfo{pages}{1891}.
\newblock


\bibitem[Monat et~al\mbox{.}(2024)]%
        {Monat24}
\bibfield{author}{\bibinfo{person}{Rapha{\"e}l Monat}, \bibinfo{person}{Aymeric Fromherz}, {and} \bibinfo{person}{Denis Merigoux}.} \bibinfo{year}{2024}\natexlab{}.
\newblock \showarticletitle{Formalizing date arithmetic and statically detecting ambiguities for the law}. In \bibinfo{booktitle}{\emph{European Symposium on Programming}}. Springer, \bibinfo{pages}{421--450}.
\newblock


\bibitem[Mora et~al\mbox{.}(2021)]%
        {Mora21}
\bibfield{author}{\bibinfo{person}{Federico Mora}, \bibinfo{person}{Murphy Berzish}, \bibinfo{person}{Mitja Kulczynski}, \bibinfo{person}{Dirk Nowotka}, {and} \bibinfo{person}{Vijay Ganesh}.} \bibinfo{year}{2021}\natexlab{}.
\newblock \showarticletitle{Z3str4: A multi-armed string solver}. In \bibinfo{booktitle}{\emph{International Symposium on Formal Methods}}. Springer, \bibinfo{pages}{389--406}.
\newblock


\bibitem[Nay et~al\mbox{.}(2024)]%
        {Nay24-LLMsTaxAttorneys}
\bibfield{author}{\bibinfo{person}{John~J Nay}, \bibinfo{person}{David Karamardian}, \bibinfo{person}{Sarah~B Lawsky}, \bibinfo{person}{Wenting Tao}, \bibinfo{person}{Meghana Bhat}, \bibinfo{person}{Raghav Jain}, \bibinfo{person}{Aaron~Travis Lee}, \bibinfo{person}{Jonathan~H Choi}, {and} \bibinfo{person}{Jungo Kasai}.} \bibinfo{year}{2024}\natexlab{}.
\newblock \showarticletitle{Large language models as tax attorneys: a case study in legal capabilities emergence}.
\newblock \bibinfo{journal}{\emph{Philosophical Transactions of the Royal Society A}} \bibinfo{volume}{382}, \bibinfo{number}{2270} (\bibinfo{year}{2024}), \bibinfo{pages}{20230159}.
\newblock


\bibitem[Novikov et~al\mbox{.}(2025)]%
        {Novikov25-AlphaEvolve}
\bibfield{author}{\bibinfo{person}{Alexander Novikov}, \bibinfo{person}{Ng{\^a}n V{\~u}}, \bibinfo{person}{Marvin Eisenberger}, \bibinfo{person}{Emilien Dupont}, \bibinfo{person}{Po-Sen Huang}, \bibinfo{person}{Adam~Zsolt Wagner}, \bibinfo{person}{Sergey Shirobokov}, \bibinfo{person}{Borislav Kozlovskii}, \bibinfo{person}{Francisco J.~R. Ruiz}, \bibinfo{person}{Abbas Mehrabian}, \bibinfo{person}{M.~Pawan Kumar}, \bibinfo{person}{Abigail See}, \bibinfo{person}{Swarat Chaudhuri}, \bibinfo{person}{George Holland}, \bibinfo{person}{Alex Davies}, \bibinfo{person}{Sebastian Nowozin}, \bibinfo{person}{Pushmeet Kohli}, {and} \bibinfo{person}{Matej Balog}.} \bibinfo{year}{2025}\natexlab{}.
\newblock \bibinfo{title}{{AlphaEvolve}: A Coding Agent for Scientific and Algorithmic Discovery}.
\newblock
\showeprint[arxiv]{2506.13131}~[cs.AI]
\urldef\tempurl%
\url{https://arxiv.org/abs/2506.13131}
\showURL{%
\tempurl}


\bibitem[Padhye(2024)]%
        {Padhye24-SEAI4Law}
\bibfield{author}{\bibinfo{person}{Rohan Padhye}.} \bibinfo{year}{2024}\natexlab{}.
\newblock \showarticletitle{Software engineering methods for {AI}-driven deductive legal reasoning}. In \bibinfo{booktitle}{\emph{Proceedings of the 2024 ACM SIGPLAN International Symposium on New Ideas, New Paradigms, and Reflections on Programming and Software}}. \bibinfo{pages}{85--95}.
\newblock


\bibitem[Pierce(2012)]%
        {BirthdayPuzzle}
\bibfield{author}{\bibinfo{person}{Rod Pierce}.} \bibinfo{year}{2012}\natexlab{}.
\newblock \bibinfo{title}{Birthday Puzzle}.
\newblock \bibinfo{howpublished}{\url{https://www.mathsisfun.com/puzzles/birthday.html}}.
\newblock
\shownote{Retrieved Jan 10, 2026}.
\newblock


\bibitem[Pimpalkhare et~al\mbox{.}(2021)]%
        {medleysolver}
\bibfield{author}{\bibinfo{person}{Nikhil Pimpalkhare}, \bibinfo{person}{Federico Mora}, \bibinfo{person}{Elizabeth Polgreen}, {and} \bibinfo{person}{Sanjit~A. Seshia}.} \bibinfo{year}{2021}\natexlab{}.
\newblock \showarticletitle{{MedleySolver}: Online {SMT} Algorithm Selection}. In \bibinfo{booktitle}{\emph{Theory and Applications of Satisfiability Testing -- {SAT} 2021}} \emph{(\bibinfo{series}{Lecture Notes in Computer Science}, Vol.~\bibinfo{volume}{12831})}. \bibinfo{publisher}{Springer}, \bibinfo{pages}{453--470}.
\newblock
\href{https://doi.org/10.1007/978-3-030-80223-3_31}{doi:\nolinkurl{10.1007/978-3-030-80223-3_31}}


\bibitem[Presburger(1929)]%
        {Presburger29}
\bibfield{author}{\bibinfo{person}{Moj\.zesz Presburger}.} \bibinfo{year}{1929}\natexlab{}.
\newblock \showarticletitle{\"{U}ber die Vollst\"{a}ndigkeit eines gewissen Systems der Arithmetik ganzer Zahlen, in welchem die Addition als einzige Operation hervortritt}. In \bibinfo{booktitle}{\emph{Comptes Rendus du I Congr\`es des Math\'ematiciens des Pays Slaves, Warszawa}}. \bibinfo{publisher}{Sk{\l}ad G{\l}\'owny}, \bibinfo{pages}{92--101}.
\newblock


\bibitem[Reingold and Dershowitz(2018)]%
        {Reingold18}
\bibfield{author}{\bibinfo{person}{Edward~M. Reingold} {and} \bibinfo{person}{Nachum Dershowitz}.} \bibinfo{year}{2018}\natexlab{}.
\newblock \bibinfo{booktitle}{\emph{Calendrical Calculations: The Ultimate Edition} (\bibinfo{edition}{4} ed.)}.
\newblock \bibinfo{publisher}{Cambridge University Press}.
\newblock


\bibitem[Reuters(2024)]%
        {Reuters24}
\bibfield{author}{\bibinfo{person}{Reuters}.} \bibinfo{year}{2024}\natexlab{}.
\newblock \bibinfo{title}{'Leap year glitch' shuts some New Zealand fuel pumps}.
\newblock \bibinfo{howpublished}{\url{https://www.reuters.com/world/asia-pacific/leap-year-glitch-shuts-some-new-zealand-fuel-pumps-2024-02-29/}}.
\newblock


\bibitem[Rice(1976)]%
        {Rice76}
\bibfield{author}{\bibinfo{person}{John~R. Rice}.} \bibinfo{year}{1976}\natexlab{}.
\newblock \showarticletitle{The Algorithm Selection Problem}.
\newblock In \bibinfo{booktitle}{\emph{Advances in Computers}}. Vol.~\bibinfo{volume}{15}. \bibinfo{publisher}{Elsevier}, \bibinfo{pages}{65--118}.
\newblock
\href{https://doi.org/10.1016/S0065-2458(08)60520-3}{doi:\nolinkurl{10.1016/S0065-2458(08)60520-3}}


\bibitem[Saxena et~al\mbox{.}(2010)]%
        {Saxena10}
\bibfield{author}{\bibinfo{person}{Prateek Saxena}, \bibinfo{person}{Devdatta Akhawe}, \bibinfo{person}{Steve Hanna}, \bibinfo{person}{Feng Mao}, \bibinfo{person}{Stephen McCamant}, {and} \bibinfo{person}{Dawn Song}.} \bibinfo{year}{2010}\natexlab{}.
\newblock \showarticletitle{A symbolic execution framework for javascript}. In \bibinfo{booktitle}{\emph{2010 IEEE Symposium on Security and Privacy}}. IEEE, \bibinfo{pages}{513--528}.
\newblock


\bibitem[Schanely(2026)]%
        {CrossHair}
\bibfield{author}{\bibinfo{person}{Phillip Schanely}.} \bibinfo{year}{2026}\natexlab{}.
\newblock \bibinfo{title}{{CrossHair}: An analysis tool for {Python} that blurs the line between testing and type systems}.
\newblock \bibinfo{howpublished}{\url{https://github.com/pschanely/CrossHair}}.
\newblock


\bibitem[Scott et~al\mbox{.}(2021)]%
        {machsmt}
\bibfield{author}{\bibinfo{person}{Joseph Scott}, \bibinfo{person}{Aina Niemetz}, \bibinfo{person}{Mathias Preiner}, \bibinfo{person}{Saeed Nejati}, {and} \bibinfo{person}{Vijay Ganesh}.} \bibinfo{year}{2021}\natexlab{}.
\newblock \showarticletitle{MachSMT: A machine learning-based algorithm selector for SMT solvers}. In \bibinfo{booktitle}{\emph{International Conference on Tools and Algorithms for the Construction and Analysis of Systems}}. Springer, \bibinfo{pages}{303--325}.
\newblock


\bibitem[Tiwari et~al\mbox{.}(2025)]%
        {Tiwari25}
\bibfield{author}{\bibinfo{person}{Shrey Tiwari}, \bibinfo{person}{Serena Chen}, \bibinfo{person}{Alexander Joukov}, \bibinfo{person}{Peter Vandervelde}, \bibinfo{person}{Ao Li}, {and} \bibinfo{person}{Rohan Padhye}.} \bibinfo{year}{2025}\natexlab{}.
\newblock \showarticletitle{It’s About Time: An Empirical Study of Date and Time Bugs in Open-Source Python Software}. In \bibinfo{booktitle}{\emph{2025 IEEE/ACM 22nd International Conference on Mining Software Repositories (MSR)}}. \bibinfo{pages}{39--51}.
\newblock
\href{https://doi.org/10.1109/MSR66628.2025.00020}{doi:\nolinkurl{10.1109/MSR66628.2025.00020}}


\bibitem[Trinh et~al\mbox{.}(2014)]%
        {Trinh14}
\bibfield{author}{\bibinfo{person}{Minh-Thai Trinh}, \bibinfo{person}{Duc-Hiep Chu}, {and} \bibinfo{person}{Joxan Jaffar}.} \bibinfo{year}{2014}\natexlab{}.
\newblock \showarticletitle{S3: A symbolic string solver for vulnerability detection in web applications}. In \bibinfo{booktitle}{\emph{Proceedings of the 2014 ACM SIGSAC Conference on Computer and Communications Security}}. \bibinfo{pages}{1232--1243}.
\newblock


\bibitem[van~der Maaten and Hinton(2008)]%
        {vandermaaten2008tsne}
\bibfield{author}{\bibinfo{person}{Laurens van~der Maaten} {and} \bibinfo{person}{Geoffrey Hinton}.} \bibinfo{year}{2008}\natexlab{}.
\newblock \showarticletitle{Visualizing Data using t-{SNE}}.
\newblock \bibinfo{journal}{\emph{Journal of Machine Learning Research}}  \bibinfo{volume}{9} (\bibinfo{year}{2008}), \bibinfo{pages}{2579--2605}.
\newblock


\bibitem[Xiong et~al\mbox{.}(2022)]%
        {Xiong22-JSSP}
\bibfield{author}{\bibinfo{person}{Hegen Xiong}, \bibinfo{person}{Shuangyuan Shi}, \bibinfo{person}{Danni Ren}, {and} \bibinfo{person}{Jinjin Hu}.} \bibinfo{year}{2022}\natexlab{}.
\newblock \showarticletitle{A survey of job shop scheduling problem: The types and models}.
\newblock \bibinfo{journal}{\emph{Computers \& Operations Research}}  \bibinfo{volume}{142} (\bibinfo{year}{2022}), \bibinfo{pages}{105731}.
\newblock


\bibitem[Xu et~al\mbox{.}(2008)]%
        {satzilla}
\bibfield{author}{\bibinfo{person}{Lin Xu}, \bibinfo{person}{Frank Hutter}, \bibinfo{person}{Holger~H. Hoos}, {and} \bibinfo{person}{Kevin Leyton-Brown}.} \bibinfo{year}{2008}\natexlab{}.
\newblock \showarticletitle{SATzilla: portfolio-based algorithm selection for SAT}.
\newblock \bibinfo{journal}{\emph{J. Artif. Int. Res.}} \bibinfo{volume}{32}, \bibinfo{number}{1} (\bibinfo{date}{June} \bibinfo{year}{2008}), \bibinfo{pages}{565–606}.
\newblock
\showISSN{1076-9757}


\bibitem[Yoshioka et~al\mbox{.}(2021)]%
        {Yoshioka21-DataAugmentation}
\bibfield{author}{\bibinfo{person}{Masaharu Yoshioka}, \bibinfo{person}{Yasuhiro Aoki}, {and} \bibinfo{person}{Youta Suzuki}.} \bibinfo{year}{2021}\natexlab{}.
\newblock \showarticletitle{{BERT}-based ensemble methods with data augmentation for legal textual entailment in {COLIEE} statute law task}. In \bibinfo{booktitle}{\emph{Proceedings of the eighteenth international conference on artificial intelligence and law}}. \bibinfo{pages}{278--284}.
\newblock


\bibitem[Zamudio~Amaya et~al\mbox{.}(2025)]%
        {fandango25}
\bibfield{author}{\bibinfo{person}{Jos\'{e}~Antonio Zamudio~Amaya}, \bibinfo{person}{Marius Smytzek}, {and} \bibinfo{person}{Andreas Zeller}.} \bibinfo{year}{2025}\natexlab{}.
\newblock \showarticletitle{FANDANGO: Evolving Language-Based Testing}.
\newblock \bibinfo{journal}{\emph{Proc. ACM Softw. Eng.}} \bibinfo{volume}{2}, \bibinfo{number}{ISSTA}, Article \bibinfo{articleno}{ISSTA040} (\bibinfo{date}{June} \bibinfo{year}{2025}), \bibinfo{numpages}{23}~pages.
\newblock
\href{https://doi.org/10.1145/3728915}{doi:\nolinkurl{10.1145/3728915}}


\bibitem[Zheng et~al\mbox{.}(2017)]%
        {Zheng17}
\bibfield{author}{\bibinfo{person}{Yunhui Zheng}, \bibinfo{person}{Vijay Ganesh}, \bibinfo{person}{Sanu Subramanian}, \bibinfo{person}{Omer Tripp}, \bibinfo{person}{Murphy Berzish}, \bibinfo{person}{Julian Dolby}, {and} \bibinfo{person}{Xiangyu Zhang}.} \bibinfo{year}{2017}\natexlab{}.
\newblock \showarticletitle{Z3str2: an efficient solver for strings, regular expressions, and length constraints}.
\newblock \bibinfo{journal}{\emph{Formal Methods in System Design}} \bibinfo{volume}{50}, \bibinfo{number}{2} (\bibinfo{year}{2017}), \bibinfo{pages}{249--288}.
\newblock


\bibitem[Zheng et~al\mbox{.}(2013)]%
        {Zheng13}
\bibfield{author}{\bibinfo{person}{Yunhui Zheng}, \bibinfo{person}{Xiangyu Zhang}, {and} \bibinfo{person}{Vijay Ganesh}.} \bibinfo{year}{2013}\natexlab{}.
\newblock \showarticletitle{Z3-str: A z3-based string solver for web application analysis}. In \bibinfo{booktitle}{\emph{Proceedings of the 2013 9th Joint Meeting on Foundations of Software Engineering}}. \bibinfo{pages}{114--124}.
\newblock


\end{thebibliography}

\end{document}